\documentclass[pra,floatfix,nofootinbib]{revtex4}

\usepackage{times,amsmath,amsfonts,amssymb,latexsym}
\usepackage{graphics,graphicx,epsf,epsfig}
\usepackage{bbm,bm,times,color}

\newtheorem{theorem}{Theorem}

\newtheorem{definition}[theorem]{Definition}
\newtheorem{example}[theorem]{Example}

\newenvironment{proof}[1][Proof]{\noindent\textbf{#1.} }{\ \rule{0.5em}{0.5em}}

\newtheorem{myassumption}{Assumption}

\newtheorem{mylemma}{Lemma}

\newtheorem{mytheorem}{Theorem}
\newtheorem{myremark}{Remark}
\newtheorem{mycorollary}{Corollary}

\newcommand{\be}{\begin{equation}}
\newcommand{\ee}{\end{equation}}
\newcommand{\ba}{\begin{eqnarray}}
\newcommand{\ea}{\end{eqnarray}}

\newcommand{\ignore}[1]{}
\def\CC{{\rm\kern.24em \vrule width.04em height1.46ex depth-.07ex
    \kern-.30em C}}
\def\P{{\rm I\kern-.25em P}}
\def\RR{{\rm
         \vrule width.04em height1.58ex depth-.0ex
         \kern-.04em R}}

\def\bbbc{{\mathchoice {\setbox0=\hbox{$\displaystyle\rm C$}\hbox{\hbox
to0pt{\kern0.4\wd0\vrule height0.9\ht0\hss}\box0}}
{\setbox0=\hbox{$\textstyle\rm C$}\hbox{\hbox
to0pt{\kern0.4\wd0\vrule height0.9\ht0\hss}\box0}}
{\setbox0=\hbox{$\scriptstyle\rm C$}\hbox{\hbox
to0pt{\kern0.4\wd0\vrule height0.9\ht0\hss}\box0}}
{\setbox0=\hbox{$\scriptscriptstyle\rm C$}\hbox{\hbox
to0pt{\kern0.4\wd0\vrule height0.9\ht0\hss}\box0}}}}
\def\bbbz{{\mathchoice {\hbox{$\sf\textstyle Z\kern-0.4em Z$}}
{\hbox{$\sf\textstyle Z\kern-0.4em Z$}}
{\hbox{$\sf\scriptstyle Z\kern-0.3em Z$}}
{\hbox{$\sf\scriptscriptstyle Z\kern-0.2em Z$}}}}

\begin{document}

\title{Adiabatic approximation with exponential accuracy for many-body systems
and quantum computation}
\author{Daniel A. Lidar$^{(1,2,3,4)}$, Ali T. Rezakhani$^{(1,4)}$, and Alioscia Hamma$^{(1,4,5,6)}$}
\affiliation{Departments of $^{(1)}$Chemistry, $^{(2)}$Electrical Engineering, and
  $^{(3)}$Physics, and $^{(4)}$Center for
Quantum Information Science \& Technology, University of Southern
California, Los Angeles, CA 90089, USA}
\affiliation{$^{(5)}$Perimeter Institute for Theoretical Physics, 31 Caroline St. N, N2L
  2Y5, Waterloo ON, Canada}
\affiliation{$^{(6)}$Massachusetts Institute of Technology,
Research Laboratory of Electronics, 77 Massachusetts Ave., Cambridge, MA 02139, USA}

\begin{abstract}
We derive a version of the adiabatic theorem that is especially suited for
applications in adiabatic quantum computation, where it is reasonable to
assume that the adiabatic interpolation between the initial and final
Hamiltonians is controllable. Assuming that the Hamiltonian is analytic in a
finite strip around the real time axis, that some number of its
time-derivatives vanish at the initial and final times, and that the target
adiabatic eigenstate is non-degenerate and separated by a gap from the rest
of the spectrum, we show that one can obtain an error between the final
adiabatic eigenstate and the actual time-evolved state which is
exponentially small in the evolution time, where this time itself scales as
the square of the norm of the time-derivative of the Hamiltonian, divided by
the cube of the minimal gap.
\end{abstract}

\maketitle

\section{Introduction}

The adiabatic approximation, with its long history \cite{Born:28,Kato:50},
has played a central role in quantum mechanics. This approximation states,
roughly, that for a system initially prepared in an eigenstate (e.g., the
ground state) $|\Phi _{0}(0)\rangle $ of a sufficiently slowly varying
Hamiltonian $h(t)$, the time evolution governed by the Schr\"{o}dinger
equation $i\frac{\partial |\psi (t)\rangle }{\partial t}=h(t)|\psi
(t)\rangle $ will approximately keep the actual state $|\psi (t)\rangle $ of
the system in the corresponding instantaneous ground state $|\Phi
_{0}(t)\rangle $ of $h(t)$, provided that there are no level crossings.
Quantitative statements of this approximation have a long history, with
rigorous results appearing only in recent years. Let $|\Phi _{j}(t)\rangle $
($j\in \{0,1,2,\ldots \}$) denote the instantaneous eigenstate of $h(t)$
with energy $e_{j}(t)$, i.e., $h(t)|\Phi _{j}(t)\rangle =e_{j}(t)|\Phi
_{j}(t)\rangle $. The simplest and one of the oldest traditional versions of
the adiabatic approximation states that the Hamiltonian must be slow with
respect to the time scale dictated by the ratio of a matrix element of the
time-derivative of the Hamiltonian to the square of the spectral gap $%
d\equiv \min_{0\leq t\leq T;j>0}[e_{j}(t)-e_{0}(t)]$ \cite{messiah}.
Namely, the fidelity between the actual final state and the adiabatic eigenstate
satisfies $|\langle \psi (T)|\Phi _{0}(T)\rangle |^{2}\geq 1-\varepsilon
^{2} $, where $\varepsilon \equiv \max_{0\leq t\leq T;j>0}[|\langle \Phi
_{j}(t)|\dot{h}(t)|\Phi _{0}(t)\rangle |/d^{2}]$.
Unfortunately this simple
criterion --- while often useful, and widely used (e.g., in Refs. \cite%
{Farhi:00,Farhi:01,Roland:02}) --- is in fact neither necessary nor
sufficient in general, as recently confirmed experimentally
\cite{du:060403}. The inadequacy of the traditional criterion is well
known and the criterion has
been replaced by rigorous general results as can be found, e.g., in Refs.~\cite%
{Teufel:book,Nenciu:93,Avron:99,Hagedorn:02,Jansen:06}, and in the presence
of noise, in Ref.~\cite{o'hara:042319}. All these rigorous results are more
severe in the gap condition than the traditional criterion, and they involve
a power of the \emph{norm} of time derivatives of the Hamiltonian, rather
than a transition matrix element. In this work we revisit the adiabatic
approximation and prove a version of the adiabatic theorem that is motivated
by recent developments in the field of quantum information science \cite%
{Nielsen:book}, namely the idea of adiabatic quantum computation (AQC) \cite%
{Farhi:00,Farhi:01}.

AQC offers a fascinating paradigm for exploiting quantum mechanics in order
to obtain a speedup for classically difficult computational problems. In AQC
one solves a computational problem by adiabatically modifying a Hamiltonian
whose initial ground state $|\Phi _{0}(0)\rangle $ encodes the input and
whose final ground state $|\Phi _{0}(T)\rangle $ encodes the output. The
time $T$ taken to reach the final ground state is the \textquotedblleft
running time\textquotedblright\ of the quantum adiabatic algorithm, which
one would like to minimize while at the same time minimizing the distance $%
\delta $ between the actual final state $|\psi (T)\rangle $ and the desired
final ground state $|\Phi _{0}(T)\rangle $. This, of course, is the subject
of the quantum adiabatic theorem, and provides the motivation for the
present paper. In quantum computation one is interested in how $T$ scales
with problem size, which is typically encoded into the system size needed to
represent the problem, e.g., the number $n$ of quantum bits
(\textquotedblleft qubits\textquotedblright ). Thus one expects both $T$ and
$\delta $ to scale with $n$. 
% {\bf 
Our goal is two-fold. First, to determine
the scaling properties of $T$ and $\delta$ with $n$. Second, to show
that $\delta$ can be made exponentially small in $T$.
%}

AQC\ was first proposed as an approach to solving optimization problems such
as satisfiability of Boolean formulas, by encoding a cost function into the
Hamiltonian \cite{Farhi:00,Farhi:01}. However, it was soon realized that
AQC\ is not limited to optimization:\ from a computational complexity
perspective AQC is equivalent in power to all other models for universal
quantum computation \cite{Aharonov:04,Siu:04,Kempe:04,Oliveira:05,MLM:06}.
Namely, AQC\ and the other models for universal quantum computation can
simulate one another with at most polynomial resource overhead. One of the
reasons AQC\ has generated much interest recently is that it has a rich
connection to well studied problems in condensed matter physics. For
example, because of the dependence of $T$ on the minimal gap, the
performance of quantum adiabatic algorithms is strongly influenced by the
type of quantum phase transition the same system would undergo in the
thermodynamic limit \cite{Latorre:04,Schutzhold:06}, thus offering an
interesting perspective on the connection between quantum information
methods and problems in condensed matter physics. AQC also appears to offer
advantages over classical simulated annealing in finding an approximate
ground state energy of a complex system \cite{zagoskin:120503}.

To date, most AQC studies have relied on the traditional version of the
adiabatic approximation. The main question we wish to address in this work
is: what is the rigorous tradeoff between a small final error $\delta $ and
the scaling of the final time $T$ with system size $n$, or any other
relevant parameter? The reason one expects a dependence on $n$ is clear from
the \textquotedblleft traditional criterion\textquotedblright :\ the gap $d$
and $\dot{h}(t)$ both depend on $n$. It has been known for some years in the
mathematical physics literature that exponential accuracy in the form $%
\delta \leq e^{-cT}$ ($c$ is some constant) is possible \cite%
{Nenciu:93,Hagedorn:02}, but the cost in terms of physical resources such as
system size $n$ has thus far not been quantified in rigorous proofs of the
adiabatic theorem, even those specifically aimed at AQC \cite%
{Jansen:06,o'hara:042319}. However, in AQC an understanding of the scaling
of the running time with respect to
problem size is of paramount importance.

It is further important to stress that in AQC\ one deals with
\textquotedblleft designer Hamiltonians\textquotedblright , which contain
controllable parameters beyond what is typically assumed in developments of
the adiabatic approximation or proofs of the adiabatic theorem. That is, one
envisions any number of \textquotedblleft control knobs\textquotedblright\
which allow one to realize the goal of transforming the initial to the final
Hamiltonian. Here we prove a version of the adiabatic theorem which focuses
on one such control knob: the number $N$ of vanishing initial and final time
derivatives of the Hamiltonian. Furthermore, we assume that the Hamiltonian
satisfies certain analyticity properties. Our theorem states, roughly, that
for such Hamiltonians, the deviation between the final state and the desired
(non-degenerate) adiabatic eigenstate can be made exponentially small in $N$,
with a running time that scales as as a polynomial in $N$ times the square of the
supremum of the norm of the time derivative of the Hamiltonian, divided by
the cube of the minimal gap. Since the scaling of $\sup_{t}\Vert \dot{h}%
\Vert $ and of the gap $d$ can be quantified in terms of the system size $n$
for situations of interest in AQC (see Section \ref{sec:norm-bounds}), this
provides an answer to the scaling question we posed above, and our result
should be particularly useful for AQC\ applications. In proving this result
we rely
on the adiabatic exponential error estimate and asymptotic
expansion due to Hagedorn and Joye \cite{Hagedorn:02}.

The structure of this paper is the following. We begin by stating our
notation, definitions, and technical assumptions in Section
\ref{sec:prelim}, and conclude this section with a statement of our
version of the
adiabatic theorem,
including some remarks. Before proving this theorem, we provide pertinent
background in Section \ref{sec:background}, including a brief summary of key
results from Ref.~\cite{Hagedorn:02}. Section \ref{sec:proof} is devoted to
the proof of the adiabatic theorem. We provide a discussion in Section \ref%
{sec:discussion}, including a comparison of our version of the adiabatic theorem to some of
the results of Refs.~\cite{Jansen:06,Schaller:06}. In the same section we
also analyze the scaling of the running time and error with system size,
give an explicit result (Corollary \ref{cor:2QPT}) for second order quantum
phase transitions, find the minimum adiabatic time $T$ by optimizing the
adiabatic interpolation and give an application of our adiabatic theorem in the open system setting
(Theorem~\ref{th:full-open}). We conclude in Section \ref{sec:conc} with
some remarks about future directions. The reader who is not interested in
the (lengthy) details of the proof of our adiabatic theorem can safely skip
from the end of Section \ref{sec:prelim} to the discussion in Section \ref{sec:discussion}.

\section{Adiabatic theorem with arbitrary accuracy}

In this section we state our adiabatic theorem. But before doing so, we
introduce the requisite notation, definitions, and assumptions.

\label{sec:prelim}

\subsection{Schr\"{o}dinger equation in dimensionless units}

Let us start with the time-dependent Schr\"{o}dinger equation
\begin{equation}
i\frac{\partial \psi }{\partial t}=h(t)\psi (t),
\end{equation}%
where we work in units of $\hbar \equiv 1$.
Define the dimensionless Hamiltonian $H$ as
\begin{equation}
H\equiv h/J,
\end{equation}%
where $J$ is an arbitrary energy unit
relative to which we shall express all other dimensional quantities.%
\footnote{%
Since $J$ is arbitrary it will drop out at the end of the analysis, when we
reintroduce dimensional units.} Let us fix the final time $T$ and define the
dimensionless rescaled time $\tau $ as
\begin{equation}
\tau \equiv t/T = \epsilon Jt.
\end{equation}%
This expresses the fact that for any given $T$ we can always choose
$J$ so that $\epsilon$, defined via
\begin{equation}
\epsilon \equiv 1/(JT),  \label{eq:eps}
\end{equation}
is a small number, which we need for the asymptotic
expansion below.
% [A10] We need to be careful here. Because, making J large translates into making energy small. For a sort of similar reason, Aharonov et al [15] (page 174) have defined "running time" as T \times \Vert H \Vert, not T alone.
%D11 Agreed, but since J cancels out in our case, I don't think we have
%to worry too much.
The Schr\"{o}dinger equation in dimensionless units now reads
\begin{equation}
i\epsilon \frac{\partial \psi }{\partial \tau }=H(\tau )\psi (\tau ).
\label{sc}
\end{equation}%
%{\bf 
From now on most of our calculations will be done in dimensionless
units.
%}

\subsection{Assumptions concerning the Hamiltonian, target state, and
initial state}

We shall need three assumptions. The first sets the stage for the family of
Hamiltonians we shall be concerned with in this work. A Hamiltonian is by
definition a self-adjoint (i.e., Hermitian) operator for real-valued times $%
\tau $, and we shall be concerned with analytic continuations of the
Hamiltonian.

\begin{figure}[bp]
\includegraphics[width=8cm,height=4.44cm]{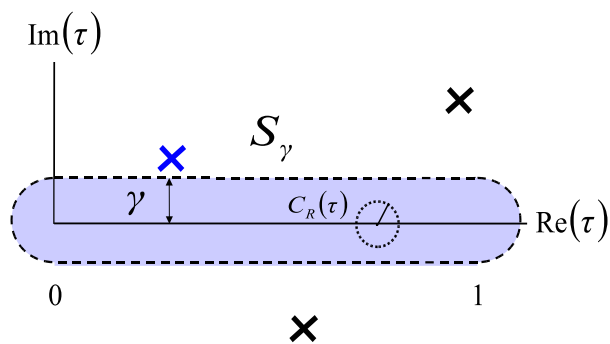}
\caption{Region of analyticity $S_{\protect\gamma }$ of the time-dependent,
analytically continued Hamiltonian $H(\protect\tau )$. The $\times $ symbols
denote possible poles or branch points of $H(\protect\tau )$. }
\label{fig-strip}
\end{figure}

\begin{myassumption}
\label{assum-1}$\{H(\tau )\}_{{\rm Re}\tau \in \lbrack 0,\infty )}$ is a
one-parameter family of bounded Hamiltonians of an $n$-body system, with the
separable Hilbert space $\mathcal{H}_{n}=\mathcal{H}^{\otimes n}$. \footnote{%
A ``separable" vector space is one that
admits a countable orthonormal basis.} Let $\gamma >0$ denote the distance
to the pole or branch point of $\{H(\tau )\}$ that is nearest to the real $%
\tau $-axis in the complex $\tau $-plane. The family $\{H(\tau
)\}_{{\rm Re}\tau \in
\lbrack 0,\infty )}$ admits an analytic continuation to an open set $%
S_{\gamma }\equiv \{\tau :|\tau |<\gamma \}\cup \{\tau :|\mathrm{\ Im}(\tau
)|<\gamma ,\mathrm{Re}(\tau )\in \lbrack 0,1]\}\cup \{\tau :|\tau -1|<\gamma
\}$, as depicted in Fig.~\ref{fig-strip}.\footnote{%
We assume analyticity in the sense of Kato \cite{Reed-Simon:book4}[p.14].
Note that if
$H(\tau)$ is a family of bounded operators, analyticity in the sense of Kato
is equivalent to the definition of bounded operator-valued analytic
functions.}
\end{myassumption}

\begin{figure}[bp]
\includegraphics[width=8cm,height=4.44cm]{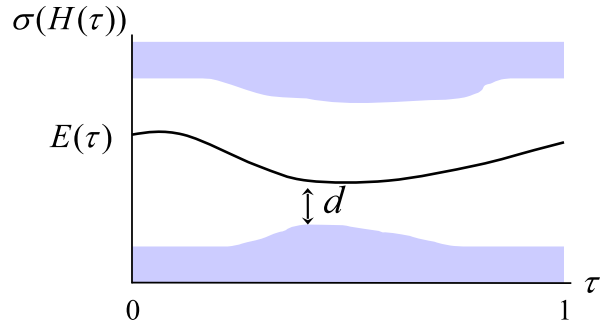}
\caption{Assumed gap structure in the spectrum of $h(\protect\tau )$.}
\label{gap}
\end{figure}

\begin{example}
In linear interpolations of the type $H(\tau )=x_0(\tau )H_{0}+x_1(\tau
)H_{1}$
(with $H_{0}$ and $H_{1}$ constant Hamiltonians such that $%
H(0)=H_{0} $ and $H(1)=H_{1}$), often used in AQC, if $x_0$ and $x_1$ are
real-analytic functions, we can usually
perform an analytic continuation. The height $\gamma$ is dictated by
possible singularities that appear because of complexification of the
functions $x_0$ and $x_1$. E.g., $x_0(\tau )=(1-\tau )/(1+\tau ^{2})$ and $%
x_1(\tau )=2\tau /(1+\tau ^{2})$ (satisfying $x_0(0)=x_1(1)=1$ and $%
x_0(1)=x_1(0)=0$), are
real-analytic
functions of $\tau \in \mathbb{R}$, but have singularities at $\tau =\pm i$.
In this case, analytical continuation is possible within $S_{\gamma }$
as depicted in Fig.~\ref{fig-strip}, with $0<\gamma <1$.
\end{example}

We denote the spectrum of an operator $A$ by $\sigma (A)$. While our
analysis applies equally well to finite and infinite-dimensional but bounded
systems ($\dim \mathcal{H}<\infty $ or $=\infty $, respectively), for most
applications the Hamiltonians of interest to us are those that are usually
considered in quantum information theory, such as spin lattices with
exchange interactions, for which $\dim \mathcal{H}<\infty $ and $\sigma
(H(\tau ))$ is discrete. Whenever our analysis below can be simplified by
this assumption, we will explicitly assume that the instantaneous
eigensystem associated with the discrete spectrum $\sigma (H(\tau ))$ is $%
\{E_{j}(\tau ),|\Phi _{j}(\tau )\rangle \}_{j=0}^{M-1}$. Therefore the
spectral theorem allows us to write:
\begin{equation}
H(\tau )=\sum_{j=0}^{M-1}E_{j}(\tau )|\Phi _{j}(\tau )\rangle \langle \Phi
_{j}(\tau )|,  \label{H-spec}
\end{equation}%
where $\langle \Phi _{i}(\tau )|\Phi _{j}(\tau )\rangle =\delta _{ij}$, and $%
M=\dim (\mathcal{H}_{n})=(\dim \mathcal{H})^{n}$. For notational convenience
we will often write $|\Phi _{0}(\tau )\rangle \equiv |\Phi (\tau )\rangle $,
and $E_{0}(\tau )\equiv E(\tau )$, though the \textquotedblleft
target\textquotedblright\ instantaneous eigenstate $|\Phi (\tau )\rangle $
need not necessarily be the ground state (see Fig.~\ref{gap}). The labeling
in Eq.~(\ref{H-spec}) is chosen such that it preserves ordering of the
eigenvalues at the initial time (eigenvalues are continuous for all real $%
\tau $, but not necessarily differentiable). This means that we allow the
eigenvalues to cross [except with $E(\tau )$] (see Assumption~\ref{assum-2}).

Our second assumption further constrains the class of Hamiltonians:

\begin{myassumption}
\label{assum-deriv}It is possible to set any given number of the derivatives of
the Hamiltonian $H(\tau )$ to zero at the initial and final times.
\end{myassumption}

This assumption is reasonable for \textquotedblleft
controllable\textquotedblright\ Hamiltonians of the type one envisions in
AQC. We shall see that it will play a key role in constraining the error and
in enforcing the right initial condition for AQC.

Our third assumption concerns the properties of the target state. Let
\begin{equation}
\text{dist}(X,Y)\equiv \inf_{x\in X,y\in Y}|x-y|,
\end{equation}
for any pair of sets $X,Y\subseteq \mathbb{C}$:

\begin{myassumption}
\label{assum-2}The \textquotedblleft target\textquotedblright\ state $|\Phi
(\tau )\rangle $, with the corresponding eigenvalue $E(\tau )$,
is a nondegenerate and isolated eigenstate of $H(\tau )$.
\end{myassumption}

\textquotedblleft Isolated\textquotedblright\ means that the spectrum $%
\sigma (H(\tau ))$ has a gap structure around $E$ --- Fig.~\ref{gap} ---
with a non-vanishing distance from the rest of the spectrum, i.e.:
\begin{equation}
\Delta _{0}(\tau )\equiv \text{dist}\left( \{E(\tau )\},\sigma (H(\tau
))\backslash \{E(\tau )\}\right) >0~\forall \tau .  \label{delta0}
\end{equation}%
The minimum dimensionless spectral gap is%
\begin{equation}
\Delta \equiv \inf_{\tau \in \lbrack 0,1]}\Delta _{0}(\tau ),
\end{equation}%
while in dimensional units we denote the minimum spectral gap by
\begin{equation}
d=J\Delta .
\end{equation}%
We shall often require the following quantity in our analysis:
\begin{equation}
\xi \equiv \sup_{\tau \in \lbrack 0,1]}\bigl\Vert \frac{dh}{d\tau }\bigr\Vert ,
\end{equation}%
where $\Vert \cdot \Vert $ denotes the standard operator norm defined in the
next subsection. Note that $\xi $ has units of energy. The projector onto
the target subspace, and its complement, will also play important roles in
our analysis:%
\begin{eqnarray}
&&P(\tau )\equiv |\Phi (\tau )\rangle \langle \Phi (\tau )|, \\
&&P_{\perp }(\tau )\equiv I-|\Phi (\tau )\rangle \langle \Phi (\tau )|.
\end{eqnarray}%
In terms of these projectors, Eq.~(\ref{H-spec}) can be replaced by the more
general representation%
\begin{equation}
H(\tau )=E(\tau )P(\tau )+P_{\perp }(\tau )H(\tau )P_{\perp }(\tau ),
\end{equation}%
which does not depend on the existence of a discrete spectrum.

\subsection{Norms and notation}

\label{norm-sect}

We use the standard operator norm defined as the maximum singular value,
which reduces for diagonalizable operators $X$ to \cite{Bhatia:book}:%
\begin{equation}
\Vert X\Vert \equiv \sup_{\Vert v\Vert =1}|\langle v|X|v\rangle |,
\end{equation}%
where $X:~\mathcal{V}\rightarrow \mathcal{V}$, $\mathcal{V}=\text{span}%
\{|m\rangle \}$ is a linear inner-product space with vectors $|v\rangle
=\sum_{m}v_{m}|m\rangle $, for which
\begin{equation}
\Vert v\Vert \equiv \sqrt{\langle v|v\rangle }=\sqrt{\sum_{m}|v_{m}|^{2}}
\label{euclid}
\end{equation}%
is the standard Euclidean norm and $\langle m|m^{\prime }\rangle =\delta
_{mm^{\prime }}$. We shall repeatedly use the submultiplicativity property
of the operator norm \cite{Bhatia:book}:%
\begin{equation}
\Vert XY\Vert \leq \Vert X\Vert \Vert Y\Vert .
\end{equation}

The minimum gap, the norm of the Hamiltonian and the rate at which the
Hamiltonian changes, scale with the number $n$ of subsystems. The parameter $%
n$ is important for AQC because it represents the size of the computational
problem. In the thermodynamic limit, systems with interesting computational
properties undergo a quantum phase transition, i.e., the system will become
gapless and the norms of the Hamiltonian will diverge. But for every finite $%
n$ the system is finite and the gap can be assumed to be finite. Thus we
introduce notation to make explicit that these quantities vary with $n$:

For all finite $n$ the quantities $1/\Delta $, $\Vert \dot{H}\Vert $, and $%
\Vert \ddot{H}\Vert $ are finite:
\begin{eqnarray}
1/\Delta &\equiv &A(n)<\infty ,  \label{D-bound} \\
\sup_{\tau \in \lbrack 0,1]}\Vert \dot{H}\Vert &\equiv &\beta (n)=\xi
(n)/J<\infty ,  \label{B-bound} \\
\sup_{\tau \in \lbrack 0,1]}\Vert \ddot{H}\Vert &\equiv &\eta (n)<\infty ,
\label{E-bound}
\end{eqnarray}%
where dot denotes $\partial /\partial \tau $. We can always take $A,\beta
,\eta >1$.

\subsection{Adiabatic Theorem}

\label{sec:AT}

We are now ready to state our main result. We wish to quantify the error in
the adiabatic approximation, i.e., the distance between the actual final
state $|\psi (T)\rangle $ and the target state $|\Phi (T)\rangle $ (both
normalized). Since in quantum mechanics $|\psi (T)\rangle $ and $|\Phi
(T)\rangle $ are rays in the projective Hilbert space $\mathcal{P}\mathcal{H}_n$
[i.e., we identify $%
e^{i\zeta }|\psi (T)\rangle $ with $|\psi (T)\rangle $ and $e^{i\chi }|\Phi
(T)\rangle $ 
with $|\Phi (T)\rangle $, where $\zeta, \chi \in \mathbb{R}$
are both arbitrary phases], the appropriate distance is $D=\arccos f$ (the
Fubini-Study distance), where $f\equiv |\langle \psi (T)|\Phi (T)\rangle |$
is the \textquotedblleft fidelity\textquotedblright . As we shall see later
[Eq.~(\ref{error-bound-interest})], we shall obtain bounds for%
\begin{equation}
\delta \equiv \Vert |\psi (T)\rangle -e^{i\chi }|\Phi (T)\rangle \Vert ,
\end{equation}%
where $\chi \in \mathbb{R}$ is a (dynamical) phase. Writing $\langle \psi
(T)|\Phi (T)\rangle = \mathcal{F}
e^{i\theta }$, where $0\leq \mathcal{f}\leq 1$ and $\theta \in
\mathbb{R}$, we have $\delta =\sqrt{2[1-\mathcal{F}\cos (\chi +\theta )]}$. Thus as
long as $\cos (\chi +\theta )>0$, an upper bound on $\delta $ is a lower
bound on $\mathcal{F}$, and hence an upper bound on $D$. From now on we shall be
concerned exclusively with the distance $\delta $.

\begin{mytheorem}
\label{th:1}Given assumptions \ref{assum-1}-\ref{assum-2}, and that the
first
%D11 $N$
$N+1$  derivatives of the Hamiltonian vanish at $\tau =0$ and $\tau =1$,
a final time $T$ which scales as
\begin{equation}
T=\frac{q}{\gamma }N\frac{\xi (n)^{2}}{d(n)^{3}},  \label{T-val}
\end{equation}%
where the \textquotedblleft time dilation\textquotedblright\ $q>1$ is a free
parameter, yields an adiabatic approximation error which satisfies:%
\begin{equation}
\delta \leq (N+1)^{\gamma +1}q^{-N}.  \label{del-val}
\end{equation}
\end{mytheorem}
% [A10] My previous remark about writing  N+1 instead of N is
% back. Because, the existing form suggests that when N=0 (no
% vanishing H-derivative) T=0! If we want to keep N instead of N+1,
% one caveat is needed.
%D11 You are right. I changed ``first $N$  derivatives'' to ``first
%$N+1$  derivatives'' above.

Thus, the adiabatic approximation error is exponentially small in the number
$N$ of vanishing $H$-derivatives, for a given time dilation $q$.

In several other papers, e.g., Refs. \cite{Nenciu:93,Hagedorn:02},
exponential error estimates in $T$ that behave as $\delta \leq e^{-cT}$ have
been proven, where $c$ is some system-dependent
constant. This applies in the setting where $T\ $ is treated as an
independent parameter and one considers a constant
(time-dependent) Hamiltonian in a fixed Hilbert space.
However, there are some important differences between our result
and the works cited above.
%D11 First, in neither of Refs.~\cite{Nenciu:93,Hagedorn:02} is it
%possible to extract the time scale from the small parameter
%$\epsilon$ used in their expansions, which means that there is no
%direct physical interpretation of these results in terms of the adiabatic time.
First, neither of Refs.~\cite{Nenciu:93,Hagedorn:02} extracted the
time scale from the small parameter $\epsilon$ used in their
expansions, which means that no direct physical interpretation of
these results was given in terms of the adiabatic time.
% [A10] We'd better be cautious not to undermine works by others. The
% wording here is unnecessarily too strong. Besides, I wouldn't say in
% neither of the Nenciu and the HJ papers "it is not possible  ..."
% (because, in principle, it is possible to obtain time scaling from
% those works.)
%D11 Agreed -- see above.
As we shall see, $\epsilon$ alone does not have a physical meaning. It
is the combination of $\epsilon$ with the dimensional Hamiltonian
and gap that will yield the time scale for the magnitude of the
error. For the same reason, $J$ drops out [see
Eq.~(\ref{delta-1})]. Second, in Ref.~\cite{Hagedorn:02}
the error is not with respect to the instantaneous
adiabatic eigenstate, but with respect to the superadiabatic basis,
% [A10] In the HJ, yes, but in the Nenciu, no.
%D11 I removed Nenciu
which has no immediate
%D11 physical meaning or
use for AQC.
% [A10] I'm not sure about this.
%D11 I removed ``physical meaning''. Indeed, Berry even proposed an
%experiment to measure in the superadiabatic basis.
Let us now recast Theorem~\ref{th:1} in the form of an exponentially
small error as a function of $T$, by eliminating $N$ between
Eqs.~(\ref{T-val})\ and (\ref{del-val}).
In doing so we focus on the (simplified) exponential part
($q^{-N}$), since for $N\gg1$ the polynomial factor
$(N+1)^{\gamma+1}$ will be suppressed in comparison to the
exponential factor.
This gives an expression involving $q^{-1/q}$,
which is minimized when $q=e$. We thus have (after re-inserting the
polynomial factor):

\begin{mycorollary}
Under the assumptions of Theorem \ref{th:1} the adiabatic error satisfies%
\begin{equation}
  \delta \lesssim \left( cT+1\right) ^{\gamma +1}e^{-cT},
\end{equation}%
where
\begin{equation}
c\equiv \frac{\gamma d^{3}}{e\xi ^{2}}.
\end{equation}
\end{mycorollary}

Thus our analysis reproduces the exponential error estimates obtained
previously, but makes the dependence on the gap and rate of change of the
Hamiltonian explicit.

Finally, in applications to AQC it may be sufficient to have a fixed error,
as long as this corresponds to a success probability greater than $1/2$. It
then follows from the Chernoff inequality \cite{Tempo:book} that one can
then boost the confidence in the correct answer by repeating the adiabatic
computation
algorithm a few times.\footnote{%
See Ref.~\cite{MohseniRezakhaniLidar} for an extensive discussion of
%D11 estimation errors in quantum algorithms.
related issues in quantum estimation algorithms.}
% [A10] To be precise, in this reference we have only addressed errors
% in quantum estimation (although more or less rules apply to
% algorithms as well, there are more issues to consider in the case of
% algorithms.)
%D11 Right -- reformulated above.
By fixing an upper bound $\delta
_{u}$ for the error, i.e., setting $\delta \leq \delta _{u}\equiv
(N+1)^{\gamma +1}q^{-N}$, we can eliminate $q$ and relate $T$ directly to $%
\delta _{u}$:

\begin{mycorollary}
Under the assumptions of Theorem \ref{th:1} a final time $T$ which scales as
\begin{equation}
T=(\delta _{u})^{-\frac{1}{N}}\frac{1}{\gamma }N(N+1)^{\frac{\gamma +1}{N}}%
\frac{\xi (n)^{2}}{d(n)^{3}},
\end{equation}%
yields an adiabatic error which satisfies:%
\begin{equation}
\delta \leq \delta _{u}.
\end{equation}
\end{mycorollary}

This corollary demonstrates that the adiabatic time $T$ is insensitive to
the adiabatic error (in that
it depends only on its $N$th root).

\begin{myremark}
The $n$-dependence of both $\xi $ and the gap $d$ is an important aspect of
our result. It will be made explicit in Section \ref{sec:norm-bounds}.  We
also note again why we are interested in the $N$-dependence of $T$ and $%
\delta $. The reason is that we view $N$ as a controllable parameter, which
an experimenter or quantum algorithm designer can vary in order to optimize $%
T$ and $\delta $ (see Section \ref{sec:norm-bounds} for more details).
\end{myremark}

\begin{myremark}
\label{rem:gamma}The value of $\gamma $ (distance of the nearest pole
of the Hamiltonian in the complex time plane from the real time axis)
may, in general, depend on the system size $n$. However, under certain natural
assumptions for AQC which we specify in Section \ref{sec:norm-bounds}, $%
\gamma $ is $n$-independent, so that the entire $n$-dependence of $T$ comes
from $\xi $ and the gap $d$.
\end{myremark}

\begin{myremark}
Because of our analyticity assumption, it is important that only a finite
number $N$ of
derivatives of the Hamiltonian vanish at the initial and final
times, for otherwise Liouville's theorem would imply that $H(\tau )$ would
have to be constant (within its analyticity domain).
\end{myremark}

Before proving Theorem~\ref{th:1} we briefly discuss resolvents and collect
the pertinent results from Ref.~\cite{Hagedorn:02} in Section \ref{sec:background}.

\section{Background}

\label{sec:background}

\subsection{Resolvents}

\begin{definition}
The (full) resolvent of $H$ is:
\begin{equation}
R(\tau ,z)\equiv \lbrack H(\tau )-z]^{-1}.  \label{full-resolvent}
\end{equation}%
It is defined on the resolvent set%
\begin{equation}
\rho (H(\tau ))\equiv \mathbb{C}-\sigma (H(\tau )).
\end{equation}
\end{definition}

The resolvent is an analytic function of $z$ on $\rho (H)$, and where $%
H(\tau )$ is an analytic function of $\tau $, $R(\tau ,z)$ will be an
analytic function of $\tau $ as well \cite{Birman:book}. By differentiating
the identity $(H(\tau )-z)R(\tau ,z)=I$, we obtain
\begin{equation}
\dot{R}(\tau ,z)=-R(\tau ,z)\dot{H}(\tau )R(\tau ,z),  \label{II-R-1}
\end{equation}%
assuming that $z$ and $\tau $ are independent.

\begin{definition}
\label{G-definition}The reduced resolvent is a map from the full Hilbert
space to the orthogonal complement of the target subspace:
\begin{equation}
G_{r}(\tau )\equiv i[H(\tau )-E(\tau )]_{r}^{-1}:~\mathcal{H}_{n}\rightarrow
\mathcal{H}_{n}^{\bot }\equiv \mathcal{R}[P_{\perp }(\tau )],
\end{equation}%
(where $\mathcal{R}$ denotes the range) and is defined via
\begin{equation}
G_{r}(\tau )[H(\tau )-E(\tau )]=[H(\tau )-E(\tau )]G_{r}(\tau )=iP_{\perp
}(\tau ).  \label{GH}
\end{equation}
\end{definition}

An explicit representation can be given in the case of a discrete spectrum
[Eq.~(\ref{H-spec})]:
\begin{equation}
G_{r}(\tau )=\sum_{j>0}\frac{1}{\Delta _{j0}(\tau )}|\Phi _{j}(\tau )\rangle
\langle \Phi _{j}(\tau)|,  \label{Gr}
\end{equation}%
where $\Delta _{j0}$ is the $j$th energy gap from the target state:
\begin{equation}
\Delta _{j0}(\tau )\equiv E_{j}(\tau )-E(\tau ).
\end{equation}%
Also note that:
\begin{equation}
G_{r}(\tau )P_{\perp }(\tau )=P_{\perp }(\tau )G_{r}(\tau )=G_{r}(\tau ).
\label{*}
\end{equation}

\subsection{Summary of main results from Ref.~\protect\cite{Hagedorn:02},
and a modification}

The main results we shall need from Ref.~\cite{Hagedorn:02} are their
Eqs.~(2.10)--(2.19), which we reproduce here for convenience. First is the
asymptotic expansion of an approximation to the full solution of the
Schr\"{o}dinger equation, where the zeroth order is the target
adiabatic eigenstate:
\begin{eqnarray}
|\Psi _{N}(\tau ,\epsilon )\rangle &=&e^{-i\int_{0}^{\tau }E(\tau ^{\prime
})d\tau ^{\prime }/\epsilon }( |\Phi (\tau )\rangle +\epsilon |\psi
_{1}(\tau )\rangle  + \epsilon ^{2}|\psi _{2}(\tau )\rangle + \ldots 
 + \epsilon
^{N}|\psi _{N}(\tau )\rangle  + \epsilon ^{N+1}|\psi _{N+1}^{\perp }(\tau
)\rangle ) .  \label{2.15}
\end{eqnarray}
The parameter $\epsilon $ is the one defined in Eq.~(\ref{eq:eps}). The
vector $|\Psi _{N}(\tau ,\epsilon )\rangle $ first appeared in Berry's work
on the \textquotedblleft superadiabatic approximation\textquotedblright\
\cite{Berry:89supad}, and is designed to provide a close approximation to
the actual time-evolved state, under adiabatic evolution. We shall call $%
|\Psi _{N}(\tau ,\epsilon )\rangle $ the \textquotedblleft superadiabatic
state\textquotedblright , but it is important to note that it is not
normalized, while $|\Phi (\tau )\rangle $ is. By inserting the
superadiabatic state into the Schr\"{o}dinger equation, and by choosing the
phase of $|\Phi (\tau )\rangle $ so that
\begin{equation}
\langle \dot{\Phi}(\tau )|\Phi (\tau )\rangle =0,
\label{****}
\end{equation}%
the following expressions have been obtained in Ref.~\cite{Hagedorn:02}:
\begin{eqnarray}  \label{2.13}
|\psi _{j}(\tau )\rangle &=&f_{j}(\tau )|\Phi (\tau )\rangle +|\psi
_{j}^{\perp }(\tau )\rangle ,~1\leq j\leq N  \label{2.10} \\
|\psi _{j}^{\perp }(\tau )\rangle &=&G_{r}(\tau )\left( f_{j-1}(\tau )|\dot{%
\Phi}(\tau )\rangle +P_{\perp }(\tau )|\dot{\psi}_{j-1}^{\perp }(\tau
)\rangle \right) ; \quad
|\psi _{0}^{\perp }(\tau )\rangle \equiv 0,
\label{2.14a}
\end{eqnarray}%
where it follows from Eq.~(\ref{*}) that
\begin{equation}
\langle \Phi (\tau )|\psi _{j}^{\perp }(\tau )\rangle =0,~1\leq j\leq N+1,
\label{eq:perp}
\end{equation}%
explaining the \textquotedblleft $\perp$\textquotedblright\ superscript.

The functions $f_{j}(\tau )$, $1\leq j\leq N$, are determined by integration
from their defining equation $\dot{f}_{j}=-\langle \Phi |\dot{\psi}%
_{j}^{\perp }\rangle $ [Eq.~(2.11) in Ref.~\cite{Hagedorn:02}] up to a
constant $c_{j}$:%
\begin{eqnarray}
f_{0}(\tau ) &\equiv &1, \\
f_{j}(\tau ) &=&-\int_{0}^{\tau }d\tau ^{\prime }\langle \Phi (\tau ^{\prime
})|\partial _{\tau ^{\prime }}\psi _{j}^{\perp }(\tau ^{\prime })\rangle
+c_{j} \notag \\
&=&\int_{0}^{\tau }d\tau ^{\prime }\langle \partial _{\tau ^{\prime
}}\Phi (\tau ^{\prime })|\psi _{j}^{\perp }(\tau ^{\prime })\rangle +c_{j},
\end{eqnarray}%
where the last equality follows from integration by parts and Eq.~(\ref%
{eq:perp}). In Ref.~\cite{Hagedorn:02} the constants $c_{j}$ were chosen to
be zero, i.e., $f_{j}(0)=0$,
$1\leq j\leq N$. For our purposes they are chosen
so that%
\begin{equation}
f_{j}(1)=0,\quad 1\leq j\leq N,  \label{eq:f(1)}
\end{equation}%
i.e., we choose $c_{j}=\int_{0}^{1}d\tau ^{\prime }\langle \Phi (\tau
^{\prime })|\partial _{\tau ^{\prime }}\psi _{j}^{\perp }(\tau ^{\prime
})\rangle $ for $1\leq j\leq N.$

Equation~(\ref{2.14a}) can also be written in either of the following forms:
\begin{eqnarray}
|\psi _{j}^{\perp }\rangle &=& G_{r}P_{\perp }\left( f_{j-1}|\dot{\Phi}\rangle
+|\dot{\psi}_{j-1}^{\perp }\rangle \right) \notag \\
&=&G_{r}\left( f_{j-1}|\dot{\Phi}%
\rangle +|\dot{\psi}_{j-1}^{\perp }\rangle \right) \notag \\
&=& G_{r}|\dot{\psi}_{j-1}^{\perp }\rangle -f_{j-1}\dot{G}%
_{r}|\Phi \rangle .  \label{2.14'}
\end{eqnarray}
The first equality is to emphasize that $G_{r}$ (and its derivatives) always
come with a $P_{\perp }$ (although $P_{\perp }$ is already included in $%
G_{r} $).
The last equality follows from Eq.~(\ref{rel-1-1}) derived below.

Further, according to Eq.~(2.19) in Ref.~\cite{Hagedorn:02} we have:%
\begin{equation}
\Vert |\psi (\tau ,\epsilon )\rangle -|\Psi _{N}(\tau ,\epsilon )\rangle
\Vert \leq A_{N}(\tau )\epsilon ^{N+1},  \label{main-HJ}
\end{equation}%
where%
\begin{equation}
A_{N}(\tau )\leq \int_{0}^{\tau }\bigl\Vert \frac{d\psi _{N+1}^{\perp }(\tau
^{\prime })}{d\tau ^{\prime }}\bigr\Vert d\tau ^{\prime }.  \label{AN-def}
\end{equation}

The initial condition used in Ref.~\cite{Hagedorn:02} is $|\psi (0,\epsilon
)\rangle =|\Psi _{N}(0,\epsilon )\rangle $. Using Eqs.~(\ref{2.15}) and (\ref%
{2.10}) we thus have $|\psi (0,\epsilon )\rangle =\sum_{j=0}^{N}\epsilon
^{j}f_{j}(0)|\Phi (\tau )\rangle +\sum_{j=1}^{N+1}\epsilon ^{j}|\psi
_{j}^{\perp }(0)\rangle $. We shall show in Section \ref{sec:init-cond} that
we can make $|\psi _{j}^{\perp }(\tau_1 )\rangle =0$
for $1\leq j\leq N+1$ by setting the first $N+1$ derivatives of $H$ to
zero at $\tau_1$,
and in particular at $\tau_1 =0$.
Thus we have%
\begin{equation}
|\psi (0,\epsilon )\rangle =\vartheta |\Phi (\tau )\rangle ,\quad \vartheta
\equiv \sum_{j=0}^{N}\epsilon ^{j}f_{j}(0),  \label{eq:init}
\end{equation}%
and since $|\psi (0,\epsilon )\rangle $ and $|\Phi (\tau )\rangle $ are both
normalized states, it follows that the initial state is the target state up
to a pure phase factor $\vartheta \in \mathbb{C}$, namely $|\vartheta |=1$.

\section{Proof of Theorem~1}

\label{sec:proof}

\subsection{Proof strategy:\ two \textquotedblleft adiabatic
distances\textquotedblright}

The central quantity of interest in the adiabatic theorem is the distance
between the exact solution $|\psi (\tau ,\epsilon )\rangle $ and the target
eigenstate $|\Phi (\tau )\rangle $, at the final time. We define this
\textquotedblleft adiabatic distance\textquotedblright\ up to a phase
(recall the discussion at the beginning of Section \ref{sec:AT}, where this
phase was denoted by $\chi $):
\begin{equation}
\delta \equiv \Vert |\psi (1,\epsilon )\rangle -e^{-\frac{i}{\epsilon }%
\int_{0}^{1}E(\tau ^{\prime })d\tau ^{\prime }}|\Phi (1)\rangle \Vert .
\label{error-bound-interest}
\end{equation}

Using the triangle inequality we have%
\begin{eqnarray}
\delta &\leq &\delta _{1}+\delta _{2} \\
\delta _{1} &\equiv &\Vert |\psi (1,\epsilon )\rangle -|\Psi _{N}(1,\epsilon
)\rangle \Vert \\
\delta _{2} &\equiv &\Vert e^{\frac{i}{\epsilon }\int_{0}^{1}E(\tau ^{\prime
})d\tau ^{\prime }}|\Psi _{N}(1,\epsilon )\rangle -|\Phi (1)\rangle \Vert .
\end{eqnarray}%
Our strategy is to bound $\delta $ by considering $\delta _{1}$ and $\delta
_{2}$ separately. We shall show that $\delta _{1}$ can be bounded by a
quantity that decreases exponentially in the asymptotic expansion order $N$.
This will require the use of the analyticity of the Hamiltonian (Assumption %
\ref{assum-1}). And, we shall show that $\delta _{2}$ can be made to vanish
by imposing that the first $N+1$ derivatives of the Hamiltonian vanish at
the final time (Assumption \ref{assum-deriv}). We start with $\delta _{2}$
and the initial condition.

\subsection{Role of boundary conditions on derivatives of the Hamiltonian}

\label{sec:init-cond}

We now show that by introducing boundary conditions on the derivatives of $H$%
, we can impose that the initial $|\psi (0)\rangle $ is the eigenstate $%
|\Phi (0)\rangle $ up to a phase, and that $\delta _{2}=0$. The technique of
imposing boundary conditions on the derivatives of $H$ is inspired by the
old work of Lennard \cite{Lenard:59} and {Garrido and Sancho
\cite{Garrido:62} (see also Nenciu's somewhat more recent work
\cite{Nenciu:81,Nenciu:89}). }

\begin{mylemma}
\label{thm} If $H^{(k)}(\tau _{1})=0$ for all $1\leq k\leq N$ and for some $%
\tau _{1}\in \lbrack 0,1]$ then
\begin{equation}
|\psi _{j}^{\perp }(\tau _{1})\rangle =0,~~j\in \{1,\ldots ,N\},
\label{eq:psi_j}
\end{equation}
\end{mylemma}

The proof is given in Appendix \ref{app:H-deriv}. Using this Lemma we have:

\begin{mycorollary}
\label{cor:init-cond}If $H^{(k)}(0)=0$ for all $1\leq k\leq N+1$ then the
initial condition is $|\psi (0,\epsilon )\rangle =\vartheta |\Phi (0)\rangle$, as in Eq.~(\ref{eq:init}).
\end{mycorollary}

\begin{proof}
Given above Eq.~(\ref{eq:init}).
\end{proof}

In AQC\ one envisions initializing the system in the ground state, which
represents the input of the computational problem. Therefore,
the last corollary confirms that we have the proper initial condition for AQC, since
Assumption \ref{assum-2} guarantees that the target state is non-degenerate,
and hence the global phase $\vartheta $ does not matter.

Moreover, we can make the error $\delta _{2}$ vanish:

\begin{mycorollary}
If $H^{(k)}(1)=0$ for all $1\leq k\leq N+1$ then $\delta _{2}=0$.
\end{mycorollary}

\begin{proof}
It follows from Eqs.~(\ref{2.15}) and (\ref{2.10}) that%
\begin{eqnarray}
e^{i\int_{0}^{1}E(\tau ^{\prime })d\tau ^{\prime }/\epsilon }|\Psi
_{N}(1,\epsilon )\rangle  &=&|\Phi (1)\rangle +\sum_{j=1}^{N}\epsilon
^{j}|\psi _{j}(1)\rangle +\epsilon ^{N+1}|\psi _{N+1}^{\perp }(1)\rangle
\notag \\
&=&|\Phi (1)\rangle +\sum_{j=1}^{N}\epsilon ^{j}f_{j}(1)|\Phi (1)\rangle
+\sum_{j=1}^{N+1}\epsilon ^{j}|\psi _{j}^{\perp }(1)\rangle
\end{eqnarray}%
Using Lemma \ref{thm} at $\tau _{1}=1$ and recalling that $f_{0}(\tau )=1$
along with the boundary conditions $f_{j}(1)=0$ for $j\geq 1$ [Eq.~(\ref%
{eq:f(1)})] we thus have%
\begin{equation}
e^{i\int_{0}^{1}E(\tau ^{\prime })d\tau ^{\prime }/\epsilon }|\Psi
_{N}(1,\epsilon )\rangle =|\Phi (1)\rangle ,
\end{equation}%
from which it follows that $\delta _{2}=0$.
\end{proof}

\subsection{Operator and state bounds}

In this section we derive bounds (many of which are not particularly tight)
on the various operators and states that arise in the proof of Theorem~\ref%
{th:1}.

From the definition (\ref{Gr}) of the reduced resolvent and the definition
of the operator norm as the maximum eigenvalue we immediately obtain
\begin{equation}
\Vert G_{r}(\tau )\Vert =\frac{1}{\text{dist}\left( \{E(\tau )\},\sigma
(H(\tau ))\backslash \{E(\tau )\}\right) }\leq 1/\Delta =A(n).
\label{G-bound}
\end{equation}

\begin{mylemma}
\label{phi-dot-phi-ddot}%
\begin{eqnarray}  \label{Phi-ddot}
|\dot{\Phi}(\tau )\rangle &=&iG_{r}(\tau )\dot{H}(\tau )|\Phi (\tau )\rangle
,  \label{Phi-dot} \\
\dot{E}(\tau ) &=&\langle \Phi (\tau )|\dot{H}(\tau )|\Phi (\tau )\rangle ,
\label{E-dot} \\
\ddot{E}(\tau ) &=&\langle \dot{\Phi}(\tau )|\dot{H}(\tau )|\Phi (\tau
)\rangle +\langle \Phi (\tau )|\dot{H}(\tau )|\dot{\Phi}(\tau )\rangle
+\langle \Phi (\tau )|\ddot{H}(\tau )|\Phi (\tau )\rangle .  \label{E-ddot}
\\
iP_{\perp }(\tau )|\ddot{\Phi}(\tau )\rangle &=&-G_{r}(\tau )\left( \ddot{H}%
(\tau )-\ddot{E}(\tau )\right) |\Phi (\tau )\rangle -2G_{r}(\tau )\left(
\dot{H}(\tau )-\dot{E}(\tau )\right) |\dot{\Phi}(\tau )\rangle ,  \notag \\
&&  \label{Pp-Phi-ddot}
\end{eqnarray}
\end{mylemma}

Equation~(\ref{E-dot}) is also known as the Hellmann-Feynman relation.

\begin{proof}
From Eq.~(\ref{GH}), we have:
\begin{eqnarray}
G_{r}(\tau )\dot{H}(\tau ) &=&i\dot{P}_{\perp }(\tau )-\dot{G}_{r}(\tau
)\left( H(\tau )-E(\tau )\right) +\dot{E}(\tau )G_{r}(\tau )~\Longrightarrow
~ \\
G_{r}(\tau )\dot{H}(\tau )|\Phi (\tau )\rangle &=&-i|\dot{\Phi}(\tau
)\rangle \langle \Phi (\tau )|\Phi (\tau )\rangle -i|\Phi (\tau )\rangle
\langle \dot{\Phi}(\tau )|\Phi (\tau )\rangle   \notag \\
&&-\dot{G}_{r}(\tau )\left(
H(\tau )-E(\tau )\right) |\Phi (\tau )\rangle +\dot{E}(\tau )G_{r}(\tau )|\Phi (\tau )\rangle .  \label{temp-1}
\end{eqnarray}%
The second term on the RHS of Eq.~(\ref{temp-1}) vanishes because of the phase condition (\ref{****}) \cite{Hagedorn:02}; the third term vanishes because $H(\tau
)-E(\tau )$ projects onto $\mathcal{H}_{n}^{\bot }$; the last term vanishes
by Eq.~(\ref{*}). Thus using Eq.~(\ref{Gr}) we obtain the following formula for $|\dot{\Phi%
}(\tau )\rangle $:
\begin{eqnarray}
|\dot{\Phi}(\tau )\rangle &=&iG_{r}(\tau )\dot{H}(\tau )|\Phi (\tau )\rangle
\\
&=&\sum_{j\neq 0}\frac{\langle \Phi _{j}(\tau )|\dot{H}(\tau )|\Phi (\tau
)\rangle }{\Delta _{j0}(\tau )}|\Phi _{j}(\tau )\rangle ,
\end{eqnarray}%
where the last equality holds for the case of a discrete spectrum, i.e.,
when Eq.~(\ref{H-spec}) applies.

To obtain the Hellmann-Feynman relation (\ref{E-dot}), we differentiate the
relations $E(\tau )=\langle \Phi (\tau )|H(\tau )|\Phi (\tau )\rangle $ and $%
\langle \Phi (\tau )|\Phi (\tau )\rangle =1$. Another differentiation yields
Eq.~(\ref{E-ddot}).

To obtain Eq.~(\ref{Pp-Phi-ddot}), we differentiate the eigenvalue equation $%
H|\Phi \rangle =E|\Phi \rangle $ twice. Thus:
\begin{equation}
(H-E)|\ddot{\Phi}\rangle =-(\ddot{H}-\ddot{E})|\Phi \rangle -2(\dot{H}-\dot{E%
})|\dot{\Phi}\rangle .
\end{equation}%
Multiplying from the left by $G_{r}$ and using Eq.~(\ref{GH}), we obtain
\begin{equation}
iP_{\perp }|\ddot{\Phi}\rangle =-G_{r}(\ddot{H}-\ddot{E})|\Phi \rangle
-2G_{r}(\dot{H}-\dot{E})|\dot{\Phi}\rangle .
\end{equation}
\end{proof}

\begin{mycorollary}
\label{bound-norms-phi-d-dd}
\begin{eqnarray}
\| \dot{\Phi}\| &\leq &A(n)\beta (n), \label{Phi-dot-bound} \\
\| \dot{P}_{\perp }\| &\leq & 2A(n)\beta (n), \label{P-dot-bound}\\
\| P_{\perp }|\ddot{\Phi}\rangle \| &\leq &6A(n)^2\beta(n)^2+2A(n)\eta
(n).  \label{Phi-ddot-bound}
\end{eqnarray}
\end{mycorollary}

\begin{proof}
Equations~(\ref{Phi-dot-bound}) and (\ref{P-dot-bound}) are immediate from Eqs.~(\ref{B-bound}), (\ref%
{G-bound}), (\ref{Phi-dot}), and $\Vert \Phi (\tau )\Vert =1$. From the
Hellmann-Feynman relation, Eq.~(\ref{E-dot}), and the definition of the
operator norm, we have: $|\dot{E}(\tau )|\leq \Vert \dot{H}(\tau )\Vert $.
Combining this with Eq.~(\ref{E-ddot}), we also obtain: $|\ddot{E}(\tau
)|\leq 2\Vert \dot{H}(\tau )\Vert \Vert \dot{\Phi}(\tau )\Vert +\Vert \ddot{H%
}(\tau )\Vert $. Inserting these relations into Eq.~(\ref{Phi-ddot}),
yields:
\begin{equation}
\Vert P_{\perp }|\ddot{\Phi}\rangle \Vert \leq \Vert G_{r}\Vert \left( \Vert
\ddot{H}\Vert +\max |\ddot{E}|+2(\Vert \dot{H}\Vert +\max |\dot{E}|)\Vert
\dot{\Phi}\Vert \right) .  \label{pp-phi-ddot}
\end{equation}%
Hence,
\begin{equation}
\Vert P_{\perp }|\ddot{\Phi}\rangle \Vert \leq \Vert G_{r}\Vert \left(
2\Vert \ddot{H}\Vert +6\Vert \dot{H}\Vert \Vert \dot{\Phi}\Vert \right)
\leq 6A^{2}(n)\beta ^{2}(n)+2A(n)\eta (n).
\end{equation}
\end{proof}

\begin{myremark}
In Section~\ref{sec:norm-bounds} we consider a general family of
Hamiltonians relevant for AQC, for which we find that $\beta (n)$ and $\eta
(n)$ have the same scaling with $n$ [Eqs.~(\ref{bound-on-H}) and (\ref%
{bound-on-dH})]. In this case we have the following bound,
after we upper bound $2A(n)\eta (n)$ by $2A^{2}(n)\beta ^{2}(n)$ (as is
always possible for large enough $n$):
\begin{equation}
\Vert P_{\perp }|\ddot{\Phi}\rangle \Vert \leq 8A^{2}(n)\beta ^{2}(n).
\label{eq:rrr}
\end{equation}
\end{myremark}

\begin{mycorollary}
\label{G-properties}%
\begin{eqnarray}
G_{r}(\tau )|\Phi (\tau )\rangle &=&\langle \Phi (\tau )|G_{r}(\tau )=0,
\label{G-Phi} \\
\langle \Phi (\tau )|\dot{G}_{r}(\tau ) &=&-\langle \dot{\Phi}(\tau
)|G_{r}(\tau ),  \label{rel-1} \\
\dot{G}_{r}(\tau )|\Phi (\tau )\rangle &=&-G_{r}(\tau )|\dot{\Phi}(\tau
)\rangle ,  \label{rel-1-1} \\
\dot{G}_{r}(\tau )P_{\perp }(\tau ) &=&\dot{P}_{\perp }(\tau )G_{r}(\tau
)+iG_{r}(\tau )[\dot{H}(\tau )-\dot{E}(\tau )]G_{r}(\tau ),  \label{eeq1} \\
\Vert \dot{G}_{r}(\tau )P_{\perp }(\tau )\Vert &\leq &4A^{2}(n)\beta (n).
\label{g-p-perp-bound}
\end{eqnarray}
\end{mycorollary}

\begin{proof}
Equation (\ref{G-Phi}) follows from Eq.~(\ref{*}). By differentiating Eq.~(%
\ref{G-Phi}) we immediately obtain Eqs.~(\ref{rel-1}) and (\ref{rel-1-1}).
From Eq.~(\ref{GH}) we have
\begin{equation}
\dot{G}_{r}(\tau )[H(\tau )-E(\tau )]=i\dot{P}_{\perp }(\tau )-G_{r}(\tau )[%
\dot{H}(\tau )-\dot{E}(\tau )],
\end{equation}%
or after multiplying from the right by $G_{r}$ and using Eq.~(\ref{GH})
again:
\begin{equation}
\dot{G}_{r}(\tau )P_{\perp }(\tau )=\dot{P}_{\perp }(\tau )G_{r}(\tau
)+iG_{r}(\tau )[\dot{H}(\tau )-\dot{E}(\tau )]G_{r}(\tau ).  \label{temp-2}
\end{equation}

Bounding the norm of $\dot{G}_{r}P_{\perp }$ suffices for our analysis,
for which Eqs.~(\ref{G-bound}), (\ref{E-dot}), (\ref{Phi-dot-bound}), and (%
\ref{temp-2}) yield:
\begin{equation}
\Vert \dot{G}_{r}(\tau )P_{\perp }(\tau )\Vert \leq 2A\beta \cdot A+A\cdot
2\beta \cdot A=4A^{2}(n)\beta (n).  \label{eq:qqq}
\end{equation}
\end{proof}

In the next lemma, and later in Eqs.~(\ref{F}) and (\ref{A-bound}), is where
we use the assumption of analyticity of the Hamiltonian. This assumption is
crucial for our error bound. The key technical tool is the Cauchy integral
formula (which requires analyticity), which links the $m$th derivative of a
function to its values in an explicit $m$-dependent way.

\begin{mylemma}
\label{3.1HJ} Let $B(0)=1$ and $B(k)=k^{k}$, and let $D(k)$ and $\widetilde{D%
}(k)$ be arbitrary functions such that $\widetilde{D}(k)\geq D(k)$ $\forall
k\in \mathbb{N}$. Suppose $\varphi (\tau )$ is an analytic vector-valued
function in the domain $S_{\gamma }$ (Fig.~\ref{fig-strip}). If $\varphi
(\tau )$ satisfies
\begin{equation}
\Vert \varphi (\tau )\Vert \leq D(k)B(k)(\gamma -|\mathrm{Im}(\tau )|)^{-k},
\label{lemma3.1-assumption}
\end{equation}%
for some $k\geq 0$, then $\dot{\varphi}(\tau )\equiv d\varphi (\tau )/d\tau $
satisfies
\begin{equation}
\Vert \dot{\varphi}(\tau )\Vert \leq \widetilde{D}(k)B(k+1)(\gamma -|\mathrm{%
Im}(\tau )|)^{-(k+1)}.  \label{3.1}
\end{equation}
\end{mylemma}

\begin{proof}
We reproduce and slightly generalize the proof reported in Ref.~\cite%
{Hagedorn:02} (Lemma 3.1). This proof is more general than what is required
in our case (Assumption \ref{assum-1}), and we include it for completeness.
Assume that $H(\tau )$ has poles or branch points at $\{\tau _{i}\}$, and
let
\begin{equation}
\mu \equiv \min_{i}\{|\mathrm{Im}(\tau _{i})|\}.
\end{equation}%
We consider a scenario wherein $H(\tau )$ can be analytically continued to
the singularity-free open set (recall Fig. \ref{fig-strip})
\begin{equation}
S_{\mu }\equiv \{\tau :|\tau |<\mu \}\cup \{\tau :|\mathrm{Im}(\tau )|<\mu ,%
\mathrm{Re}(\tau )\in \lbrack 0,1]\}\cup \{\tau :|\tau -1|<\mu \}.
\end{equation}%
Define the circle $C_{R}(\tau )=\{\tau ^{\prime }\in S_{\mu }:~|\tau -\tau
^{\prime }|=R(\tau )>0\}$, centered at $\tau $ and with radius $R(\tau )$.
This circle will shortly serve as an integration contour, and to ensure that
$C_{R}(\tau )$ is always inside the set $S_{\mu }$, it suffices to choose $%
\max_{\tau ^{\prime }\in C_{R}}|\text{Im}(\tau ^{\prime })|=|\text{Im}(\tau
)+R(\tau )|<\mu $. Taking, for example, $R(\tau )=\frac{\mu -|\text{Im}(\tau
)|}{k+1}$, for some $k\geq 0$, satisfies this requirement. Relative to
Assumption \ref{assum-1}, where $\tau =1$ and $\mu =\gamma $, all we require
is $R<\gamma $.

The main idea is to use the Cauchy integral formula for the analytic
function $\varphi (\tau )$ to write
\begin{equation*}
\dot{\varphi}(\tau )=\frac{1}{2\pi i}\oint_{C_{R}(\tau )}\frac{\varphi (\tau
^{\prime })}{(\tau ^{\prime}-\tau)^2}d\tau ^{\prime },
\end{equation*}%
where the circle $C_{R}(\tau )$ has radius $R(\tau )=\frac{\gamma -|\text{Im}%
(\tau )|}{k+1}$. For $\tau ^{\prime }\in C_{R}(\tau )$, we have $|\text{Im}%
(\tau ^{\prime })|\leq |\text{Im}(\tau )|+R(\tau )$, hence $\gamma -|\text{Im%
}(\tau ^{\prime })|\geq \frac{k}{k+1}(\gamma -|\text{Im}(\tau )|)$.
Replacing this bound in $\| \varphi (\tau )\| \leq D(k)B(k)(\gamma -|$Im$%
(\tau )|)^{-k}$ [Eq.~(\ref{lemma3.1-assumption})], we obtain
\begin{equation}
\| \varphi (\tau ^{\prime })\| \leq D(k)\left( \frac{k}{(\gamma -|\text{Im}%
(\tau ^{\prime })|)}\right) ^{k}\leq D(k)\left( \frac{k}{\frac{k}{k+1}%
[\gamma -|\text{Im}(\tau )|]}\right) ^{k}.
\end{equation}%
Let $\widetilde{D}(k)$ be any function that upper-bounds $D(k)$ for all $k$.
Then this gives rise to
\begin{eqnarray}
\| \dot{\varphi}(\tau )\| &=&\frac{1}{2\pi }\| \oint_{C_{R}(\tau )}\frac{%
\varphi (\tau ^{\prime })}{(\tau ^{\prime}-\tau)^2}d\tau ^{\prime }\|  \notag
\\
&\leq &\frac{1}{2\pi }\cdot 2\pi \frac{\gamma -|\text{Im}(\tau )|}{k+1}\cdot
\widetilde{D}(k)\cdot\left( \frac{k}{\frac{k}{k+1}[\gamma -|\text{Im}(\tau
)|]}\right) ^{k}\cdot \left( \frac{1}{k+1}[\gamma -|\text{Im}(\tau
)|]\right) ^{-2}  \notag \\
&=&\widetilde{D}(k)\left( \frac{k+1}{\gamma -|\text{Im}(\tau )|}\right)
^{k+1}.  \label{phidot}
\end{eqnarray}%
The case $k=0$ follows from
the same argument by replacing $C_{R}(\tau )$
with $\alpha (\gamma -|\text{Im}(\tau )|)$, for an arbitrary $\alpha <1$.
This results in the bound $\| \dot{\varphi}(\tau )\| \leq \widetilde{D}%
^{-1}(\gamma -|\text{Im}(\tau )|)^{-1}$. In our application of Lemma~\ref%
{3.1HJ} we use $D(k)=C(k)A^{a(k)}\beta ^{b(k)}$ and $\widetilde{D}(k)
=C(k)A^{c(k)} \beta^{d(k)}$, where $c(k)\geq a(k)$ and $d(k)\geq b(k)$, and
where $C(k)$ is given in Eq.~(\ref{C(N)}), $a(k)$ and $b(k)$ are the
functions defined later
in Eq.~(\ref{ab}), $c(k)$ and $d(k)$ are the functions
defined later
in Eq.~(\ref{cd}), and $A$ and $\beta $ are defined in Eqs.~(\ref%
{D-bound}) and (\ref{B-bound}) (both are $>1$), respectively.
\end{proof}

\subsection{Bound on $\protect\delta_1$}

The error term $\delta _{1}$ is exactly the one which already appeared in
Eq.~(\ref{AN-def}). Our strategy is to bound the integral in Eq.~(\ref%
{AN-def}) by using an inductive approach based on Lemma~\ref{3.1HJ}. So far
we have only used the differentiability property of $H(\tau )$, not its
analyticity. If one wants to assume only that $H(\tau )$ is $N+1$-times
differentiable, it is still possible to find an upper bound for the integral
in Eq.~(\ref{AN-def}), of course in terms of $\Delta $ as well as norms of
derivatives of $H(\tau )$ --- for an analysis based only on
differentiability of the Hamiltonian, see, for example, Refs.~\cite{Jansen:06,Nenciu:93}.

To apply Lemma~\ref{3.1HJ}, we first need to justify why $|\psi
_{N+1}^{\perp }(\tau )\rangle $ is analytic in $S_{\gamma }$. We first show
that by Assumption~\ref{assum-1}, $E(\tau )$
and $|\Phi (\tau )\rangle $
are analytic functions inside $S_{\gamma }$. To do so we recall the
following theorem (modified slightly for our purpose here) \cite%
{Reed-Simon:book4}:

\begin{mytheorem}[Kato-Rellich theorem]
Let $Q(\tau )$ be a family of bounded operator-valued analytic functions in
a region $S$. Let $q(\tau _{1})$ be a nondegenerate eigenvalue of $Q(\tau
_{1})$ --- for our purpose, we take $\tau _{1}\in \mathbb{R}$. Then, for $%
\tau $ near $\tau _{1}$, there is exactly one point $q(\tau )$ of $\sigma
(Q(\tau ))$ near $q(\tau _{1})$ and this point is isolated and
nondegenerate. $q(\tau )$ is an analytic function of $\tau $ near $\tau _{1}$%
, and there is an analytic eigenvector $|q(\tau )\rangle $. When $Q(\tau )$
is self-adjoint for $\tau -\tau _{1}\in \mathbb{R}$, $|q(\tau )\rangle $ is
also normalizable.
\end{mytheorem}

\begin{mycorollary}
$|\Phi (\tau )\rangle $ and $E(\tau )$ are analytic inside $S_{\gamma }$.
\end{mycorollary}

\begin{mytheorem}[XII.7 \protect\cite{Reed-Simon:book4}]
Let $Q(\tau )$ be a family of bounded operator-valued analytic functions in
a region $S$. Then the resolvent $R(\tau ,z)$ of $Q$, for $z\in \rho (Q(\tau
))$, is an analytic function of $\tau $ in $S$.
\end{mytheorem}

\begin{mycorollary}
\label{G-analytic}$G_{r}(\tau )$ is an analytic function of $\tau $ inside $%
S_{\gamma }$.
\end{mycorollary}

\begin{proof}
Let us take an arbitrary $z\in \rho (H(\tau ))$. Then, by multiplying Eq.~(%
\ref{GH}) by $R(\tau ,z)$ from the right, we obtain
\begin{eqnarray*}
&&G_{r}(\tau )\left[ (H(\tau )-z)-(E(\tau )-z)\right] =iP_{\perp }(\tau
)~\Longrightarrow \\
&&G_{r}(\tau )=\frac{i}{z-E(\tau )}P_{\perp }(\tau )R(\tau ,z)\left[
P_{\perp }(\tau )R(\tau ,z)-\frac{I}{E(\tau )-z}\right] ^{-1}.
\end{eqnarray*}%
Note that, since $\sigma (P_{\perp }(\tau )R(\tau ,z))=\{\frac{1}{E(\tau )-z}%
;~E(\tau )\in \sigma (H(\tau ))-\{E(\tau )\},z\in \rho (H(\tau ))\}$, the
inverse on the RHS exists. Moreover, the expression on the RHS is analytic
in terms of $R(\tau )$, which, together with analyticity of $P_{\perp }(\tau
)$ in $\tau \in S_{\gamma }$, implies analyticity of $G_{r}(\tau )$ in $%
S_{\gamma }$.
\end{proof}

\begin{mycorollary}
\label{an-psi}$|\psi _{N+1}^{\perp }(\tau )\rangle $ is analytic inside $%
S_{\gamma }$.
\end{mycorollary}

\begin{proof}
Note that $|\psi _{1}^{\perp }(\tau )\rangle =G_{r}(\tau )|\dot{\Phi}(\tau
)\rangle $ is analytic, because of analyticity of $G_{r}(\tau )$ and $|\Phi
(\tau )\rangle $ in $S_{\gamma }$, and because differentiation preserves
analyticity. Now, let us assume by induction that $|\psi _{j}^{\perp }(\tau
)\rangle $ is analytic. Analyticity of $|\psi _{j+1}^{\perp }(\tau )\rangle $
is then immediate from Eqs.~(\ref{2.13}) and (\ref{2.14'}):
\begin{equation*}
|\psi _{j+1}^{\perp }(\tau )\rangle =G_{r}(\tau )\left[ |\dot{\Phi}(\tau
)\rangle \int_{0}^{\tau }d\tau ^{\prime }\langle \dot{\Phi}(\tau ^{\prime
})|\psi _{j}^{\perp }(\tau )\rangle+|\dot{\psi}_{j}^{\perp }(\tau )\rangle %
\right] ,\quad 1\leq j\leq N,
\end{equation*}%
and the RHS is a product and sum of analytic functions.
\end{proof}

Now we can initiate the application of Lemma~\ref{3.1HJ}. We assume that $%
\tau \in \mathbb{R}$ since we are only concerned with real time evolution in
AQC.
For notational simplicity we drop the $\tau $-dependence from here on. From
Eq.~(\ref{2.14'}) we have:
\begin{equation}
|\psi _{N}^{\bot }\rangle =G_{r}P_{\perp }\left[ f_{N-1}|\dot{\Phi}\rangle +|%
\dot{\psi}_{N-1}^{\bot }\rangle \right] .  \label{2.14}
\end{equation}

We will find $\Vert \dot{\psi}_{N}^{\bot }\Vert $ by induction. To
initialize the induction we use Eqs.~(\ref{2.14a}) and (\ref{2.13}):%
\begin{eqnarray}
|\psi _{1}^{\bot }\rangle &=&G_{r}P_{\perp }|\dot{\Phi}\rangle \quad
\overset{(\ref{G-bound}),(\ref{Phi-dot-bound})}{\Longrightarrow }\quad \Vert
\psi _{1}^{\bot }\Vert \leq A^{2}\beta ,  \label{1} \\
\overset{ (\ref{G-bound}),(\ref{Phi-dot-bound}),(\ref{P-dot-bound}),(\ref{Phi-ddot-bound}),(%
\ref{g-p-perp-bound})}{\Longrightarrow }|\dot{\psi}_{1}^{\bot }\rangle &=&\dot{%
G}_{r}P_{\perp }|\dot{\Phi}\rangle +G_{r}\dot{P}_{\perp }|\dot{\Phi}\rangle
+G_{r}P_{\perp }|\ddot{\Phi}\rangle  \notag \\
\overset{(\ref{eq:rrr}),(\ref{eq:qqq})}{\Longrightarrow }\quad \Vert
\dot{\psi}_{1}^{\bot }\Vert &\leq& 14A^{3}\beta ^{2}.  \label{2}
\end{eqnarray}
We now assume by induction that
\begin{equation}
\Vert \psi _{N}^{\bot }\Vert \leq C(N)g(N)A^{a(N)}\beta ^{b(N)},
\label{psiperp}
\end{equation}
where $a(N)$, $b(N)$ and $C(N)$ are functions we shall determine, and where
\begin{equation}
g(N)=\left( \frac{N-1}{\gamma }\right) ^{N-1}\quad \lbrack g(1)\equiv 1].
\label{f}
\end{equation}%
The form of $g(N)$ comes from Lemma~\ref{3.1HJ}. \footnote{%
Note that we have set $\mathrm{Im}(\tau )=0$ because we are on the real axis.%
} Lemma~\ref{3.1HJ} determines that differentiation raises $g(N)$ to $g(N+1)$%
. Setting $D(N)=C(N)A^{a(N)}\beta ^{b(N)}$ and $\widetilde{D}%
=C(N)A^{c(N)}\beta ^{d(N)}$ in that Lemma, provided $c(N)\geq a(N)$ and $%
d(N)\geq b(N)$, we obtain the bound:
\begin{equation}
\Vert \dot{\psi}_{N}^{\bot }\Vert \leq C(N)g(N+1)A^{c(N)}\beta ^{d(N)}.
\label{psidot}
\end{equation}%
We shall determine $c(N)$ and $d(N)$ below.

Note that the initial conditions determined by Eqs.~(\ref{1}) and (\ref{2})
are: \footnote{%
Note that the $g(2)$ factor present in Eq.~(\ref{psidot}) (evaluated at $N=1$
and absent in Eq.~(\ref{2})) gives rise to a discrepancy between the two
bounds, unless we set $\gamma=1/14$.
This does in fact not impose a
constraint on the family of Hamiltonians our proof applies to (recall
Assumption \ref{assum-1}), since in the application of Cauchy's theorem
we are free to choose an arbitrarily small integration contour around the
real-time axis. In spite of having thus fixed its value, we continue to
write $\gamma$ rather than $1/14$,
as there is no fundamental importance to
this value; it is merely an outcome of our rather loose bounds, e.g., as in
Eq.~(\ref{eq:rrr}).}
\begin{align}
a(1)& =2\quad b(1)=1,  \label{init1} \\
c(1)& =3\quad d(1)=2,  \label{init2} \\
C(1)& =1.  \label{init3}
\end{align}

We now use the induction hypothesis to write%
\begin{equation}
\Vert \psi _{N+1}^{\bot }\Vert \leq C(N+1)g(N+1)A^{a(N+1)}\beta ^{b(N+1)},
\label{ind}
\end{equation}%
while on the other hand we have from Eq.~(\ref{2.14})%
\begin{align}
\Vert \psi _{N+1}^{\bot }\Vert & =\Vert G_{r}\bigl[ f_{N}|\dot{\Phi}\rangle
+|\dot{\psi}_{N}^{\bot }\rangle \bigr] \Vert  \notag \\
& \leq \Vert G_{r}\Vert \bigl[ |f_{N}|\Vert \dot{\Phi}\Vert +\Vert \dot{\psi}%
_{N}^{\bot }\Vert \bigr] .  \label{N+1}
\end{align}%
The inductive proof consists of showing that the bounds (\ref{ind}) and (\ref%
{N+1}) are the same. To do so we first need to bound $\left\vert
f_{N}\right\vert $:%
\begin{align}
\left\vert f_{N}(\tau )\right\vert & =\left\vert \int_{0}^{\tau }\langle
\dot{\Phi}(\tau ^{\prime })|\psi _{N}^{\bot }(\tau ^{\prime })\rangle d\tau
^{\prime }\right\vert  \notag \\
& \leq \int_{0}^{\tau }\left\vert \langle \dot{\Phi}(\tau ^{\prime })|\psi
_{N}^{\bot }(\tau ^{\prime })\rangle \right\vert d\tau ^{\prime }  \notag \\
& \leq \tau \cdot \sup_{\tau ^{\prime} \in [0,\tau]}\Vert \dot{\Phi}(\tau
^{\prime })\Vert \Vert \psi _{N}^{\bot }(\tau ^{\prime })\Vert  \notag \\
& \leq 1\cdot A\beta \cdot C(N)g(N)A^{a(N)}\beta ^{b(N)},  \label{F}
\end{align}%
where in the last step we used the induction hypothesis again, and hence
also the analyticity assumption. Using this bound in (\ref{N+1}) together
with (\ref{psidot}) we find:%
\begin{align}
\Vert \psi _{N+1}^{\bot }\Vert & \leq A\left[ C(N)g(N)A^{a(N)+1}\beta
^{b(N)+1}\cdot A\beta +1\cdot C(N)g(N+1)A^{c(N)}\beta ^{d(N)}\right]  \notag
\\
& =C(N)\left[ g(N)A^{a(N)+3}\beta ^{b(N)+2}+g(N+1)A^{c(N)+1}\beta ^{d(N)}%
\right]  \notag \\
& \leq C(N)\left[ g(N)A^{a(N)+3}\beta ^{b(N)+2}+g(N+1)A^{c(N)+1}\beta
^{d(N)}(A\beta )^{k}\right] .  \label{N+1:2}
\end{align}%
In the last inequality we multiplied the second term by $(A\beta )^{k}$,
where $k\geq 1$ is a constant, in order to allow for an adjustment to fit
the initial conditions; see below. In order to complete the inductive proof
the two bounds (\ref{ind})\ and (\ref{N+1:2}) should agree, and for this it
is sufficient that their RHSs are equal:%
\begin{eqnarray}
C(N+1)g(N+1)A^{a(N+1)}\beta ^{b(N+1)}=C(N)\left[ g(N)A^{a(N)+3}\beta
  ^{b(N)+2} +g(N+1)A^{c(N)+1+k}\beta ^{d(N)+k}\right] .
\end{eqnarray}%
Since $A$ and $\beta $ are arbitrary this requires that
\begin{equation}
C(N+1)g(N+1)=C(N)\left[ g(N)+g(N+1)\right] ,  \label{Cf}
\end{equation}%
while the terms involving $A$ and $\beta $ must have equal powers, which
implies:%
\begin{align}
a(N+1)& =a(N)+3=c(N)+1+k, \\
b(N+1)& =b(N)+2=d(N)+k.
\end{align}%
From these last two equations, together with the initial conditions (\ref%
{init1}) we easily find:%
\begin{equation}
a(N)=3N-1,\quad b(N)=2N-1.  \label{ab}
\end{equation}%
We also have $c(N)=3N+1-k$ and $d(N)=2N+1-k$. The initial conditions (\ref%
{init2}) then yield $k=1$, so that: \footnote{%
Another way to understand the need for the adjustment in the last line of
Eq.~(\ref{N+1:2}) comes from this example:
\begin{align*}
\Vert \psi _{2}^{\bot }\Vert & \leq \Vert G_{r}\Vert \left[ |f_{1}|\Vert
\dot{\Phi}\Vert +\Vert P_{\bot }\Vert \Vert \dot{\psi}_{1}^{\bot }\Vert %
\right] \\
& \leq A^{5}\beta ^{3}+A^{4}\beta ^{2}\leq 2A^{5}\beta ^{3}.
\end{align*}%
To get the last inequality we multiplied the term $A^{4}\beta ^{2}$ by $%
(A\beta )^{k}$ with $k=1$. This is required in order to obtain a bound
involving just a single power of $A$ and of $\beta $. Failing to do this
allows for the possibility that the two bounds (\ref{ind}) and (\ref{N+1:2})
will not agree.}
\begin{equation}
c(N)=3N,\quad d(N)=2N.  \label{cd}
\end{equation}

Next we need to solve for $C(N)$ from Eq.~(\ref{Cf}), subject to the initial
condition $C(1)=14$.
We have, using Eq.~(\ref{f}):%
\begin{equation}
C(N+1)=C(N)\left( 1+\gamma \frac{\left( N-1\right) ^{N-1}}{N^{N}}\right) ,
\label{C-rec}
\end{equation}%
whose solution is%
\begin{equation}
C(N)=\prod_{j=1}^{N-1}\left( 1+\gamma \frac{\left( j-1\right) ^{j-1}}{j^{j}}%
\right) .  \label{C(N)}
\end{equation}

We can upper-bound $C(N)$ as follows:
\begin{equation}
C(N)\leq \prod_{j=1}^{N-1}\left( 1+\gamma \frac{j^{j-1}}{j^{j}}\right)
=\prod_{j=1}^{N-1}\frac{j+\gamma }{j}\leq \prod_{j=1}^{N-1}\frac{j+m}{j},
\end{equation}%
where $m=\left\lceil \gamma \right\rceil $ (smallest integer larger than $%
\gamma $). Thus%
\begin{eqnarray}
C(N) &\leq &\prod_{j=1}^{N-1}\frac{j+m}{j}=\frac{N(N+1)\cdots (N-1+m)}{m!} \\
&\leq &(N+1)(\frac{N}{2}+1)\cdots (\frac{N}{m}+1)  \notag \\
&\leq &(N+1)^{m}\leq (N+1)^{\gamma +1}.
\end{eqnarray}%

By collecting all our results and inserting them into Eq.~(\ref{psidot}) we
have, so far:%
\begin{equation}
\Vert \dot{\psi}_{N}^{\bot }\Vert \leq (N+1)^{\gamma +1}\left( \frac{%
NA^{3}\beta ^{2}}{\gamma }\right) ^{N}.
\end{equation}%
From Eq.~(\ref{AN-def}), we now have:%
\begin{equation}
A_{N}(\tau )\leq \tau \cdot \sup_{0\leq s\leq \tau \leq 1}\Vert \dot{\psi}%
_{N+1}^{\bot }(s)\Vert \leq (N+2)^{\gamma +1}\left( \frac{(N+1)A^{3}\beta
^{2}}{\gamma }\right) ^{N+1}.  \label{A-bound}
\end{equation}%
Reinserting dimensional units (i.e., $H=h/J$ and $\Delta =d/J$) we have
\begin{align}
\delta _{1}& \equiv \Vert |\psi (1,\epsilon )\rangle -|\Psi _{N}(1,\epsilon
)\rangle \Vert \leq A_{N}(1)\epsilon ^{N+1}  \notag \\
& \leq (N+2)^{\gamma +1}\left( \frac{(N+1)(d/J)^{-3}(\sup_{\tau \in \lbrack
0,1]}\Vert \dot{h}/J\Vert )^{2}}{\gamma }\right) ^{N+1}(JT)^{-(N+1)}  \notag
\\
& =(N+2)^{\gamma +1}\left( \frac{(N+1)\xi ^{2}}{\gamma Td^{3}}\right) ^{N+1},
\label{delta-1}
\end{align}%
where $\dot{h}\equiv \frac{dh}{d\tau }$.

Thus picking $T$ as
\begin{equation}
T=\frac{q}{\gamma }(N+1)\frac{\xi ^{2}}{d^{3}},
\end{equation}%
where $q>1$ is a \textquotedblleft time dilation factor\textquotedblright ,
gives:%
\begin{equation}
\delta \leq (N+2)^{\gamma +1}q^{-(N+1)}
\end{equation}%
which can be made arbitrarily small in the number $N+1$
of zero derivatives of the Hamiltonian. We have thus proved Theorem
\ref{th:1}
(where $N+1$ is redefined as $N$).
% [A10] My earlier caveat about the case N=0: when N=0 then T=0. (A remark is in order.)
%D11 Was already addressed here since N+1 is called the number of zero derivatives.

\section{Discussion}

\label{sec:discussion}

\subsection{Comparison to the results of Jensen, Ruskai, and Seiler
\protect\cite{Jansen:06}}

\label{sec:review}

In Ref.~\cite{Jansen:06}, Jansen, Ruskai, and Seiler (JRS) proved --- using
the methodology of Avron \textit{et al.} \cite{Avron:99} rather than the
asymptotic expansion \cite{Hagedorn:02} --- a number of adiabatic theorems,
all of which made weaker assumptions than ours. In particular, they did not
assume analyticity. For example, their Theorem 3 can be summarized as
follows. Assuming that $h(\tau )$ is $C^{2}$, that $\Vert \dot{h}\Vert $ and
$\Vert \ddot{h}\Vert $ are both bounded, and that $\dot{h}(0)=\dot{h}(1)=0$,
then provided
\begin{equation}
T=q\int_{0}^{1}\left( m\frac{\Vert \ddot{h}\Vert }{d_{0}^{2}}+7m\sqrt{m}%
\frac{\Vert \dot{h}\Vert ^{2}}{d_{0}^{3}}\right) d\tau ,  \label{eq:Ruskai}
\end{equation}%
the error at the final time can be made arbitrarily small in the
\textquotedblleft time dilation factor\textquotedblright\ $q>1$:%
\begin{equation}
\delta \leq q^{-2}.  \label{eq:d-ad}
\end{equation}%
Here $d_{0}(\tau )=J\Delta _{0}(\tau )$ is the instantaneous dimensional
minimal gap [Eq.~(\ref{delta0})], and the parameter $m(\tau )$ is the number
of distinct eigenvalues in the spectrum of $h$ restricted to the target
subspace.

It is interesting to compare this result to our Theorem~\ref{th:1}. In our
case, by assumption the target subspace is one-dimensional, so $m\equiv 1$
(though this does not appear to be fundamental to our analysis).
Furthermore, we have in various places replaced integrals over $\tau $ by
the supremum of their integrand. In light of this, we would have written
Eq.~(\ref{eq:Ruskai}) as
\begin{equation}
T=7q\frac{\xi ^{2}}{d^{3}},  \label{T-Ruskai}
\end{equation}%
which is indeed very similar to Eq.~(\ref{T-val}), except that the $N$%
-dependence is now absent. The reason for this is, of course, that JRS did
not consider the case of $N$
vanishing time derivatives. The fact that our
error bound (\ref{del-val})
is much tighter than JRS's (\ref{eq:d-ad}) is
again due to analyticity, which allowed us to introduce the $N$ parameter
into the error bound.

\subsection{System-size dependent bounds for local Hamiltonians}

\label{sec:norm-bounds}

Due to the appearance of $\sup_{\tau \in \lbrack 0,1]}\Vert \frac{dh}{d\tau }%
\Vert ^{2}$ in the expression for the adiabatic time (\ref{T-val}), we now
present norm bounds for physically relevant Hamiltonians, with the purpose
of exhibiting the explicit system-size dependence to the extent possible.
This is particularly relevant for AQC.

Let us consider the
case when $h(\tau )$ is an $L$-local $n$-body Hamiltonian ($L\leq n$). An $L$%
-local Hamiltonian contains interaction terms involving at most $L$ bodies,
for some fixed $L$. For example, in the $n$-qubit case let $\{\bm{\sigma }\}$
be the operator basis constructed from tensor products of the Pauli matrices
$\sigma _{x}$, $\sigma _{y}$, $\sigma _{z}$, and the identity matrix $I$.
Then we can expand the Hamiltonian as:
\begin{equation}
H(\tau )=\sum_{\bm{\sigma }}\xi _{\bm{\sigma }}(\tau )\bm{\sigma },
\label{H-inter}
\end{equation}%
in which all $\xi _{\bm{\sigma }}$ are real functions. This expression for $%
H(\tau )$ is in fact a very general \textquotedblleft
interpolation\textquotedblright\ Hamiltonian, which captures many of the
examples considered in the AQC literature, such as the common linear
interpolation Hamiltonians of the type $H(\tau )=(1-\xi (\tau ))H_{0}+\xi
(\tau )H_{1}$, where $H_{0}$ and $H_{1}$ are fixed, $n$-qubit Hamiltonians,
and $\xi (0)=0$, $\xi (1)=1$ \cite%
{Farhi:01,Roland:02,Latorre:04,Schutzhold:06,zagoskin:120503}. It also
captures the unitary interpolation $H(\tau )=U(\tau )H(0)U^{\dag }(\tau )$
\cite{Siu:04}, where $U(\tau )$ is unitary; this can be seen by Taylor
expansion of $U(\tau )$.

For an $L$-local Hamiltonian, by definition $\xi _{\bm{\sigma }}=0$ whenever
the (Hamming) weight of the corresponding $\bm{\sigma }$ --- the number of
non-identity terms in the tensor product --- is greater than $L$. The number
of independent real parameters is \cite{Haselgrove:03}
\begin{equation}
\#(n,L)=\sum_{j=0}^{L}\binom{n}{j}3^{j}\overset{L\leq n/2}{\leq }(L+1)\binom{%
n}{L}3^{L}.
\end{equation}%
In most physically relevant systems $L=2$, i.e.,
\begin{equation}
H(\tau )=\sum_{j=1}^{n}V_{j}(\tau )+\sum_{j<j^{\prime }}V_{jj^{\prime
}}(\tau ).
\end{equation}%
In such cases, we have
\begin{equation}
\#(n,2)=(9n^{2}-3n+2)/2.
\end{equation}%
Putting together all these elements we find
\begin{equation}
\Vert H(\tau )\Vert \leq \#(n,2)\cdot \sup_{\bm{\sigma}}|\xi _{\bm{\sigma}%
}(\tau )|.
\end{equation}%
Therefore, the overall upper bound for the scaling in terms of $n$ is as
follows:
\begin{equation}
\Vert H(\tau )\Vert =O\left( n^{2}\cdot \sup_{\bm{\sigma}}|\xi _{\bm{\sigma }%
}(\tau )|\right) \leq O\left( n^{2}\cdot \sup_{\bm{\sigma},\tau }|\xi _{%
\bm{\sigma }}(\tau )|\right) .
\end{equation}%
With a similar analysis, we also obtain
\begin{equation}
\Vert H^{(k)}(\tau )\Vert \leq O\left( n^{2}\cdot \sup_{\bm{\sigma},\tau
}|\xi _{\bm{\sigma }}^{(k)}(\tau )|\right) ,~k\in \mathbb{N},
\end{equation}%
where $(\cdot )^{(k)}\equiv \frac{\partial ^{k}}{\partial \tau
  ^{k}}(\cdot )$.
Usually, in
physical systems of interest in condensed matter or quantum
information, there is a predefined lattice or graph structure that dictates
the spatial configuration of the system. Typically, there are $n$ spins
arranged on a 1D, 2D, or 3D lattice or graph. Increasing $n$ in such systems
means adding new particles only to the surface or boundary of the lattice.
That is, by construction, a new particle cannot occupy a position inside the
lattice, unless a structural defect is present. Even if there are structural
defects in the system, it is plausible to assume that the number of
defective sites is very small relative to the total number of the particles
in the system, or this number may be a constant independent of $n$. Assuming
that the interactions are only \emph{local} ($2$-local in space and local in
time) and \emph{short-ranged}, therefore, new particles would not change the
coupling strengths between the particles far enough from the surface.
\footnote{%
Note that $\xi _{\bm{\sigma}}(\tau )$ may in general depend on $n$. One can
see this through a simple example. Imagine a cylinder of gaseous particles
with short-ranged interactions. Any particle will interact with all
particles inside a sphere of radius $r_{\text{int}}$ --- the range of the
interaction --- around it. If we add new particles to the cylinder, at some
point (i.e., at some $n$) all the space inside the shell will be occupied
(close-packed), hence, the new particles cannot interact with the particle
in the center. For such particles, the coupling strength of the interaction
with the particle in the center is effectively zero.} This argument implies
that in systems with a time-independent \footnote{%
This condition is designed to exclude a folding of the system lattice, such
as protein folding in the case of polymers or DNA molecules.} and
non-defective spatial graph/lattice structure, and short-ranged, local
interactions, the coupling strengths (or interpolation functions) $\xi _{%
\bm{\sigma }}(\tau )$ do not depend on the system size $n$, for large enough
$n$, and we have:
\begin{eqnarray}
&\Vert H(\tau )\Vert \leq&\sup_{\bm{\sigma},\tau }|\xi _{\bm{\sigma }}(\tau
)|\cdot O(n^{2}),  \label{bound-on-H} \\
&\Vert H^{(k)}(\tau )\Vert \leq&\sup_{\bm{\sigma},\tau }|\xi _{\bm{\sigma }%
}^{(k)}(\tau )|\cdot O(n^{2}).  \label{bound-on-dH}
\end{eqnarray}%
That is, the norm of $H(\tau )$ and all of its derivatives scale as $%
O(n^{2}) $, but with different coefficients.

A nice feature of the interpolation Hamiltonian (\ref{H-inter}) is that its
analyticity is determined entirely by the scalar interpolation functions $%
\{\xi _{\bm{\sigma }}(\tau )\}_{\bm{\sigma }}$. Let $\gamma _{\bm{\sigma }}$
denote the height of the analyticity domain of $\xi _{\bm{\sigma
}}(1)$. Then
\begin{equation}
\gamma =\inf_{\bm{\sigma }}\gamma _{\bm{\sigma }}  \label{gamma}
\end{equation}%
is the height of the analyticity domain appearing in Theorem~\ref{th:1}.
Clearly, if the interpolation functions $\xi _{\bm{\sigma }}(\tau )$ do not
depend on the system size $n$ because the system is confined to a graph with
time-independent geometry and topology, nor will $\gamma $ (recall Remark %
\ref{rem:gamma}).

The next issue concerns the scaling of the gap with $n$. As pointed out,
e.g., in Ref.~\cite{Schutzhold:06}, how the gap scales depends on whether
one is dealing with a first or higher order quantum phase transition (QPT).
First order QPTs are typically associated with exponentially small gaps,
while higher order QPTs are associated with polynomially small gaps (in both
cases, as a function of $n$). Consider first the latter, i.e.,
\begin{equation}
d(n)\sim Jn^{-z},  \label{d-scaling}
\end{equation}%
where $z>0$ is the \textquotedblleft dynamical critical
exponent\textquotedblright\ \cite{Sachdev:book}. We can now apply these
considerations to Theorem~\ref{th:1}, and find:

\begin{mycorollary}
\label{cor:2QPT}Under the same assumptions as in Theorem~\ref{th:1}, and
assuming a second order quantum phase transition [i.e., a gap that scales as
in Eq.~(\ref{d-scaling})], a time scaling as
\begin{equation}
T=\frac{q}{\gamma }N\frac{\left( \sup_{\bm{\sigma },\tau }|\dot{\xi}_{%
\bm{\sigma }}(\tau )|\right) ^{2}}{J^{3}}n^{4-3z},  \label{T-n}
\end{equation}%
yields an adiabatic approximation error which satisfies $\delta \leq
(N+1)^{\gamma +1}q^{-N)}$ (independently of $n$).
\end{mycorollary}

A different situation arises in the context of the adiabatic version of
Grover's problem \cite{Roland:02}, which is an example of a first order QPT.
In Ref.~\cite{Jansen:06}, condition (\ref{eq:Ruskai}) was applied in this
setting, where the Hamiltonian has the following form:
\begin{equation}
H_{G}(\tau )=(1-x(\tau ))(I-|\phi \rangle \langle \phi |)+x(\tau
)(I-|m\rangle \langle m|),
\end{equation}%
(where $|\phi \rangle =\sum_{i=0}^{2^{n}-1}|i\rangle /\sqrt{2^{n}}$ and $%
m\in \{0,\ldots ,2^{n}-1\}$) and the time-dependent spectral gap $d_{0}$ is
found to have the following dependence on the number of qubits $n$:
\begin{equation}
d_{0}(n,\tau )=J\sqrt{2^{-n}+4(1-2^{-n})(x(\tau )-1/2)^{2}}
\end{equation}%
The minimum gap is encountered at the critical point, where the gap scales
as $\Delta (n)=O(2^{-n/2})$. For this problem, condition (\ref{eq:Ruskai})
gives $T=O(d^{-1})$ for constant error. It is important in deriving this
result that the function $x(\tau )$ is smooth ($C^{\infty }$). This result
is much more appealing in terms of its $n$-dependence than the general
estimate (\ref{eq:Ruskai}), but it relies on the fact that the norm of the
Hamiltonian does not scale with $n$: $\Vert H_{G}\Vert \leq 1+2\sup_{\tau
}|x(\tau )|$, which is not the generic case. In the setting of our Theorem~%
\ref{th:1}, it is clear that we would find $T=O(d^{-1})$ not for constant
error, but for an error that can be made arbitrarily small. The assumption
of a smooth Hamiltonian is of course compatible with our assumption of
analyticity.

Recently, Ref.~\cite{Schaller:06} derived another error estimate for AQC
that relies on smooth interpolation and results in the same estimate for the
running time: $T=O(d^{-1})$. To obtain this result Ref.~\cite{Schaller:06}
assumed again that the norm of the Hamiltonian is bounded above by a
constant. Reference \cite{Schaller:06} also considered the case of a
constant gap and highly degenerate first excited state (i.e., a Hamiltonian
whose norm depends on $n$), and argued numerically that for a smooth
interpolation it is possible to obtain an exponential error estimate: $%
\delta =O(n\exp (-Td))$, whence a running time $T=O(d^{-1}\ln n)$ suffices
for arbitrarily small error. This case, though, is again non-generic. The
generic situation is one in which the Hamiltonian couples all the states in
the spectrum, and the spectral gap closes with $n$.

\subsection{AQC related issues}

\label{sec:AQC-comm}

In this subsection we collect a couple of
observations related to the relevance of Theorem~\ref{th:1} to AQC.

First, as a general rule, an advantage in performing AQC over classical
computation can only be guaranteed if one has \textit{a priori} knowledge of
the final time $T$, which is presumably shorter than the time required for
the execution of the corresponding classical algorithm. Similarly, in our
setting, in order to be able to set the final-time derivatives of the
Hamiltonian equal to zero, one needs to know the final time $T$. Thus both
in the general AQC setting and in our case one would like to know the gap $d$
(as well as the other, more easily computable quantities appearing in
Theorem~\ref{th:1}). While this is sometimes amenable to an analytical
solution, it is in almost all cases a very difficult problem. Fortunately,
in AQC the gap is known exactly if one starts from a quantum algorithm given
in the circuit model, by mapping this algorithm to the adiabatic model. In
such a case the gap is an easily computable function of the number of gates
\cite{Aharonov:04,Deift:06}. However, this result relies on a physically
unreasonable Hamiltonian containing $5$-body interactions, and recent
results which map such Hamiltonians to physically reasonable $2$-body
interactions (e.g., Refs.~\cite{Kempe:04,Oliveira:05}), use so called
\textquotedblleft perturbative gadgets\textquotedblright , which involve an
approximation wherein the exact expression for the gap is lost. If one does
not know the final time $T$ exactly, one can still attempt to compute an
estimate $T_{e}$ for $T$, and set the final-time derivatives of the
Hamiltonian equal to zero at $T_{e}$. Provided $T_{e}>T$, stretching the
adiabatic evolution in such a manner cannot result in a worse error than
promised by Theorem~\ref{th:1} for $T$.

Second, in the context of AQC one would like to measure the final state, in
order to extract the answer to the computation. For this reason it makes
sense to simply make the Hamiltonian constant for $t$ starting from a
value slightly smaller than $T$ --- which implies that the derivatives
of the Hamiltonian at the final time $T$ vanish.
This, of course, automatically satisfies the requirement of vanishing
final-time derivatives. One cannot make the same argument about
initialization of the computation, however: initialization is a dynamic
process (e.g., cooling into the ground state), so that one cannot keep the
Hamiltonian constant for all $t<0$.

\subsection{The open system case}

Suppose that our quantum system of interest $S$ is coupled to another system
$B$, which acts as an environment or \textquotedblleft
bath\textquotedblright . This is the setting of open quantum systems \cite%
{Breuer:book}. Together, system and bath are described by the Hamiltonian%
\begin{equation}
h(t)=h_{S}(t)\otimes I_{B}+h_{SB}+I_{S}\otimes h_{B},  \label{h-tot}
\end{equation}
where $h_{S}$, $h_{SB}$, and $h_{B}$ are the system, system-bath, and bath
Hamiltonians, respectively, and $I$ is the identity operator. The joint
Hilbert space is $\mathcal{H}=\mathcal{H}_{S}\otimes \mathcal{H}_{B}$, where
$\mathcal{H}_{S}$ and $\mathcal{H}_{B}$ are the system and bath Hilbert
spaces. Then $h_{SB}:\mathcal{H}\mapsto \mathcal{H}$, $h_{S}:\mathcal{H}%
_{S}\mapsto \mathcal{H}_{S}$, and $h_{B}:\mathcal{H}_{B}\mapsto \mathcal{H}%
_{B}$. In our analysis above we only considered $h_{S}(t)$, which implements
the adiabatic system evolution in the present case. We assume that $h_{SB}$
and $h_{B}$ are time-independent Hamiltonians. This is a reasonable physical
assumption in many cases \cite{Breuer:book}.

Coupling of the system to the bath introduces decoherence, and modifies the
adiabatic condition relative to the closed system case we have discussed
thus far \cite{SarandyLidar:04,SarandyLidar:05,Thunstrom:05,joye-2006}. We
are interested in the adiabatic theorem which describes the system state
alone. To this end, we need an appropriate distance measure. The trace
distance is defined as $D[\rho _{1},\rho _{2}]\equiv \frac{1}{2}\| \rho
_{1}-\rho _{2}\| _{1}$, where $\| A\| _{1}\equiv \mathrm{Tr}|A|$, $|A|\equiv
\sqrt{A^{\dag }A}$, and is a good distance measure between states (or
density matrices) $\rho _{1}$ and $\rho _{2}$ acting on the same Hilbert
space \cite{Nielsen:book}. A useful fact is that taking the partial trace
can only decrease the distance between states \cite{Nielsen:book}, i.e., if $%
\rho _{1} $ and $\rho _{2}$ are states in the joint system-bath Hilbert
space $\mathcal{H}$, then
\begin{equation}
D[\mathrm{Tr}_{B}\rho _{1},\mathrm{Tr}_{B}\rho _{2}]\leq D[\rho _{1},\rho
_{2}],  \label{DTrB}
\end{equation}%
where $\mathrm{Tr}_{B}$ is the partial trace operation over the bath Hilbert
space: $\mathrm{Tr}_{B}[|s\rangle \langle s^{\prime }|\otimes |b\rangle
\langle b^{\prime }|]\equiv \langle b^{\prime }|b\rangle |s\rangle \langle
s^{\prime }|$, for arbitrary states $|s\rangle ,|s^{\prime }\rangle \in
\mathcal{H}_{S}$ and $|b\rangle ,|b^{\prime }\rangle \in \mathcal{H}_{B}$.
Inequality (\ref{DTrB}) can be understood intuitively as a consequence of
the fact that by erasing information (taking the partial trace) one cannot
make states more distinguishable, i.e., their distance cannot increase.

Consider first the \emph{uncoupled} setting $h_{SB}=0$, which we denote by
the superscript $0$. The target adiabatic system state is $\rho _{S,\mathrm{%
ad}}^{0}(t)=|\Phi (t)\rangle \langle \Phi (t)|$. Let $\rho _{\mathrm{ad}%
}^{0}(t)\equiv \rho _{S,\mathrm{ad}}^{0}(t)\otimes \rho _{B}^{0}(t)$ denote
the \textquotedblleft target adiabatic joint state,\textquotedblright\ with $%
\rho _{B}^{0}(t)=e^{-ih_{B}t}\rho _{B}^{0}(0)e^{ih_{B}t}$. Let $\rho
(0)$ denote the initial joint system-bath state. The \emph{actual} state
whose time evolution is generated by $h(t)$ [Eq.~(\ref{h-tot})] is $\rho
(t)=U(t)\rho (0)U(t)^{\dag }$, where $U(t)=\mathcal{T}e^{-i\int_{0}^{t}h(t^{\prime })dt^{\prime }}$ is the propagator of the joint
system-bath dynamics, with $\mathcal{T}$ denoting time ordering. The actual
time evolved system state is $\rho _{S}(t)=\mathrm{Tr}_{B}\rho (t)$. Using
Eq.~(\ref{DTrB}), we have the following inequality:%
\begin{equation}
\delta _{S}\equiv D[\rho _{S}(T),\rho _{S,\mathrm{ad}}^{0}(T)]\leq D[\rho
(T),\rho _{\mathrm{ad}}^{0}(T)]\equiv \delta _{SB}.  \label{delS}
\end{equation}%
\emph{The distance }$\delta _{S}$\emph{\ is the distance of interest}:\ it
is the distance between the actual system state and target system adiabatic
state. The last inequality shows that it is upper-bounded by the distance $%
\delta _{SB}$ between two \textquotedblleft closed-system\textquotedblright\
states, where closed refers here to the joint system-bath entity. Because of
this, we already know the form of the adiabatic theorem for $\delta _{SB}$.
This is just Theorem~\ref{th:1}
again, with $h$ as prescribed in Eq.~(\ref%
{h-tot}). It follows from Eq.~(\ref{delS}) that we can use this upper bound
on $\delta _{SB}$ to bound $\delta _{S}$ as well. To be explicit, let us
state the theorem we thus obtain for the open system case:

\begin{mytheorem}
\label{th:full-open} Let $d$ denote the minimum gap of the full Hamiltonian $%
h(t)$ in Eq.~(\ref{h-tot}). Given assumptions \ref{assum-1}-\ref{assum-2} on
$h(t)$, assuming $h_{B}$ and $h_{SB}$ are time-independent, and that the
first $N+1$ derivatives of the Hamiltonian vanish at $\tau =0$ and $\tau =1$,a
final time $T$ which scales as
\begin{equation}
T=\frac{q}{\gamma }N\frac{\sup_{\tau \in \lbrack 0,1]}\Vert \dot{h}_{S}\Vert
^{2}}{d^{3}},
\end{equation}%
where $q>1$ is a free parameter, yields an adiabatic approximation error
which satisfies:
\begin{equation}
\delta _{S}\leq \delta _{SB}\leq (N+1)^{\gamma +1}q^{-N}
\end{equation}
\end{mytheorem}

\begin{myremark}
There is an important difference between the closed and open system cases:
the minimum gap $d$ in the open system case is the gap for the full
system-bath Hamiltonian (\ref{h-tot}), which can be expected to be
significantly smaller than for the isolated system, since generally, due to
its much larger number of degrees of freedom, the bath will introduce many
intermediate levels inside the gap depicted in Fig.~\ref{gap} for the
isolated system. This means that $T$ can be expected to be very much larger
in the open system case than for the same system without coupling to a bath.
See also Ref.~\cite{SarandyLidar:04} for a different approach leading to the
same conclusion.
\end{myremark}

\section{Conclusions}

\label{sec:conc}

In this work we aimed to bridge a gap between rigorous formulations of the
quantum adiabatic theorem, leading to exponentially tight error estimates,
and the field of adiabatic quantum computation (AQC), where knowledge of the
way the energy gap enters the conditions for the adiabatic approximation is
crucial. To this end we have presented a version of the quantum adiabatic
approximation that is useful for AQC, where there is a single non-degenerate
ground state, the number of subsystems $n$ is variable, and where the
interpolation from the initial to the final Hamiltonian is fully
controllable, at least in principle. In this case, we have shown that for a
total time $T$ scaling as the product of the cube of the inverse gap and the
square of the operator norm of $\dot{h}$, the error in the adiabatic
approximation can be made exponentially small. Since our version of the
quantum adiabatic theorem explicitly accounts for the system size dependence
(see, e.g., Corollary \ref{cor:2QPT}), this represents an advance over
previous adiabatic theorems, where either the approximation error or the
system-size dependence is not nailed down.

Our results imply that as long as our key assumption of analyticity of the
interpolation in a domain can be satisfied, along with a degree of control
that allows setting the initial and final time derivatives of the
Hamiltonian equal to zero, then from a closed-system perspective AQC has an
important fault tolerance advantage over the circuit model of quantum
computation \cite{Nielsen:book}. Namely, whereas in the circuit model even
unitary deviations from a prescribed set of gates can ruin a quantum
algorithm, in AQC large deviations are permissible, as long as the
interpolation ends at the desired final Hamiltonian, whose ground state
encodes the answer to the computational problem one is trying to solve. Of
course, this should not be misinterpreted as a claim that AQC is fully fault
tolerant. It is well known that AQC is vulnerable to interactions with the
environment \cite%
{SarandyLidar:05,Childs:01,RolandCerf:04,ashhab:052330,Tiersch:07,Amin:07},
and only preliminary steps have been taken towards a theory of fault
tolerant AQC in an open systems setting \cite{Lidar:AQC-DD,Jordan:05}. We
have also reported a corollary regarding the adiabatic theorem for open
quantum systems (Theorem~\ref{th:full-open}), which shows that the
time-scale for adiabaticity is determined by the gap of the full system-bath
Hamiltonian.

There are indications that the adiabatic approximation fails for
Hamiltonians with several independent time-scales \cite{Marzlin:04a} (see
also Ref.~\cite{Jansen:06}). This presents an interesting problem for AQC,
even in our setting of a closed system with analytic Hamiltonians. For
example, consider a situation where there is some smooth control noise on
the interpolation functions $\xi _{\bm{\sigma
}}(\tau )$, which has an independent time-scale. Then merely slowing down
the evolution by elongating $T$ will have no impact on this noise, so that
in its presence the time dilation-based error bound (\ref{del-val}) cannot
be expected to apply. In other words, noise with an intrinsic time scale
that cannot be stretched in the sense that makes $\epsilon $ in the
asymptotic expansion (\ref{2.15}) small, generates a violation of the
assumptions used to derive the adiabatic theorem. Future work on
fault-tolerant AQC should address this problem.

\begin{acknowledgments}

We are particularly grateful to Alain Joye for important discussions
and clarifications regarding Ref.~\cite{Hagedorn:02}. We would also
like to thank Mary Beth Ruskai, Ben Reichardt, and Zhaohui Wei for
helpful correspondence, and Robert Raussendorf for comments on the
problem that arises from control noise. D.A.L.'s work was sponsored by
the National Science Foundation under grants No. CCF-0726439 and No.
PHY-0803304. A.T.R. acknowledges the support of the USC\ Center for
Quantum Information Science \& Technology. A.H. acknowledges the
financial support of the Foundational Questions Institute (fqxi.org),
and a grant from xQIT at MIT. Research at Perimeter Institute for
Theoretical Physics is supported in part by the Government of Canada
through NSERC and by the Province of Ontario through MRI.
\end{acknowledgments}

\appendix

\section{Proof of Lemma~1}
\label{app:H-deriv}

We first need the following Lemma:

\begin{mylemma}
\label{H-dot-dot} If $H^{(k)}(\tau _{1})=0$ for all $1\leq k\leq N$ and for
some $\tau _{1}\in \lbrack 0,1]$, then
\begin{equation}
E^{(k)}(\tau _{1})=P^{(k)}(\tau _{1})=|\Phi ^{(k)}(\tau _{1})\rangle
=0,~k\in \{1,\ldots ,N\}.
\end{equation}
\end{mylemma}

\begin{proof}
It is well known that for a closed operator $H(\tau )$ with $E(\tau )$ an
isolated point of $\sigma (H(\tau ))$ (i.e., suppose that for some $%
\varepsilon $: $\sigma (H)\cup \{E^{\prime }:~|E-E^{\prime }|<\varepsilon
\}=\{E\}$), the corresponding projection can be written as
\begin{equation}
P(\tau )=\frac{1}{2\pi i}\oint_{|E-E^{\prime }|=r}dE^{\prime }R(\tau
,E^{\prime }),  \label{P-general}
\end{equation}%
for any $r\in (0,\varepsilon )$ \cite{Reed-Simon:book4}, where $R(\tau
,E^{\prime })$ is the full resolvent of $H(\tau )$. Since $E(\tau )$ is the
eigenvalue of $H(\tau )$ associated with the eigenstate $|\Phi (\tau
)\rangle $ we have:
\begin{equation}
E(\tau )=\text{Tr}[H(\tau )P(\tau )].  \label{E-general}
\end{equation}%
By differentiating Eqs.~(\ref{P-general}) and (\ref{E-general}) and using
Eq.~(\ref{II-R-1}), we obtain
\begin{eqnarray}
\dot{P}(\tau ) &=&-\frac{1}{2\pi i}\oint_{|E-E^{\prime }|=r}dE^{\prime
}R(\tau ,E^{\prime })\dot{H}(\tau )R(\tau ,E^{\prime }),  \label{P-d-general}
\\
\dot{E}(\tau ) &=&\text{Tr}[\dot{H}(\tau )P(\tau )]+\text{Tr}[H(\tau )\dot{P}%
(\tau )].  \label{E-d-general}
\end{eqnarray}%
Thus making $\dot{H}(\tau _{1})=0$ implies $\dot{P}(\tau _{1})=0$, which in
turn, from Eq.~(\ref{E-d-general}), implies $\dot{E}(\tau _{1})=0$ as well.
Note that $\dot{P}(\tau _{1})=0$ also implies $\dot{P}_{\perp }(\tau _{1})=0$%
. Moreover, Eq.~(\ref{Phi-dot}) yields:
\begin{equation}
|\dot{\Phi}(\tau _{1})\rangle =iG_{r}(\tau _{1})\dot{H}(\tau _{1})|\Phi
(\tau _{1})\rangle =0.
\end{equation}%
By simple applications of the Leibniz rule ($(X\cdot Y)^{(k)}=\sum_{i=0}^{k}%
\binom{k}{i}X^{(i)}\cdot Y^{(k-i)}$, for any pair of (differentiable)
objects), it can be seen that these
conclusions hold also for higher order derivatives. We show this explicitly
for $|\Phi ^{(k)}(\tau _{1})\rangle $. We obtain
\begin{equation}
|\Phi ^{(k)}(\tau _{1})\rangle =i\sum_{i=0}^{k-1}\binom{k}{i}%
G_{r}^{(i)}(\tau _{1})\left( \dot{H}|\Phi \rangle \right) ^{(k-i)}|_{\tau
_{1}}.
\end{equation}%
All terms within the summation include a derivative of $H$. The highest
derivative in the RHS is $H^{(k)}$. This is zero for all $k\leq N$. That is,
$|\Phi ^{(k)}(\tau _{1})\rangle =0$,~ $k\in \{1,\ldots ,N\}$.
\end{proof}

\begin{mycorollary}
\label{G-dot-0} Under the assumptions of Lemma~\ref{H-dot-dot}, we have
\begin{equation}
H^{(k)}(\tau _{1})=0\Longrightarrow G_{r}^{(k)}(\tau _{1})=0,~k\in
\{1,\ldots ,N\}.
\end{equation}%
\label{cor4}
\end{mycorollary}

\begin{proof}
Using Eqs.~(\ref{rel-1-1}) and the definition of $P_{\perp }$ we have $\dot{G%
}_{r}P_{\perp }=\dot{G}_{r}+G_{r}|\dot{\Phi}\rangle \langle \Phi |$. Thus,
using Eqs.~(\ref{eeq1}) and Lemma~\ref{H-dot-dot} we find:
\begin{equation*}
\dot{G}_{r}(\tau _{1})=-G_{r}(\tau _{1})|\dot{\Phi}(\tau _{1})\rangle
\langle \Phi (\tau _{1})|+\dot{P}_{\perp }(\tau _{1})G_{r}(\tau
_{1})+iG_{r}(\tau _{1})[\dot{H}(\tau _{1})-\dot{E}(\tau _{1})]G_{r}(\tau
_{1})=0.
\end{equation*}%
From the Leibniz rule we obtain
\begin{eqnarray}
G_{r}^{(k)}(\tau _{1}) &=&-\sum_{i=0}^{k-1}\binom{k-1}{i}G_{r}^{(i)}(\tau
_{1})\left( |\dot{\Phi}\rangle \langle \Phi |\right) ^{(k-1-i)}|_{\tau
  _{1}} \notag \\
&&+\sum_{i=0}^{k-1}\binom{k-1}{i}P_{\perp }^{(i+1)}(\tau
_{1})G_{r}^{(k-1-i)}(\tau _{1}) \notag \\
&&+i\sum_{i=0}^{k-1}\binom{k-1}{i}G_{r}^{(i)}(\tau _{1})\left( [\dot{H}-\dot{%
E}]G_{r}\right) ^{(k-1-i)}|_{\tau _{1}}.
\end{eqnarray}%
From Lemma~\ref{H-dot-dot}, all the terms within the summations vanish for
all $k\leq N$.
\end{proof}

We are now ready to give the proof of Lemma~\ref{thm}.

\begin{proof}[of Lemma~\protect\ref{thm}]
We note that $|\psi _{{1}}^{\perp }(\tau )\rangle =G_{r}(\tau )|\dot{%
\Phi}(\tau )\rangle \overset{\text{(\ref{rel-1-1})}}{=}-\dot{G}_{r}(\tau
)|\Phi (\tau )\rangle $, from which by using Corollary \ref{G-dot-0} we
obtain
\begin{eqnarray}
&|\psi _{1}^{\perp }(\tau _{1})\rangle =&-\dot{G}_{r}(\tau _{1})|\Phi (\tau
_{1})\rangle =0,  \label{p-1-p-0} \\
&|\dot{\psi}_{1}^{\perp }(\tau _{1})\rangle =&-\ddot{G}_{r}(\tau _{1})|\Phi
(\tau _{1})\rangle -\dot{G}_{r}(\tau _{1})|\dot{\Phi}(\tau _{1})\rangle =0,
\label{p-1-p-d-0}
\end{eqnarray}%
or in general, using the Leibniz rule:
\begin{eqnarray}
\partial _{\tau }^{k}|\psi _{1}^{\perp }(\tau _{1})\rangle =-\sum_{i=0}^{k}%
\binom{k}{i}G_{r}^{(i+1)}(\tau _{1})|\Phi ^{(k-i)}(\tau _{1})\rangle
=0~,0\leq k\leq N-1,  \label{p-1-p-d}
\end{eqnarray}
since each term inside the summation vanishes as long as $G_{r}^{(i+1)}(1)=0$%
, i.e., $i+1\leq N$, or, $k=N-1$.

We now show by induction that if $\partial _{\tau }^{k}|\psi _{j-1}^{\perp
}(\tau _{1})\rangle =0$ for $0\leq k\leq N-(j-1)$, then $\partial _{\tau
}^{k}|\psi _{j}^{\perp }(\tau _{1})\rangle =0$ for $0\leq k\leq N-j$. The
calculation above initialized the induction for $j=2$. We have by Eq.~(\ref%
{2.14'}):
\begin{eqnarray}
\partial _{\tau }^{k}|\psi _{j}^{\perp }(\tau _{1})\rangle &=&-f_{j-1}(\tau
_{1})\left( \dot{G}_{r}|\Phi \rangle \right) ^{(k)}-\sum_{i=1}^{k}\binom{k}{i%
}f_{j-1}^{(i)}(\tau _{1})\left( \dot{G}_{r}|\Phi \rangle \right)
^{(k-i)}  \notag \\
&&+\sum_{i=1}^{k}\binom{k}{i}G_{r}^{(i)}(\tau _{1})|\psi
_{j-1}^{(k-i+1)}(\tau _{1})\rangle +G_{r}(\tau _{1})|\psi _{j-1}^{(k+1)}(\tau _{1})\rangle .
\end{eqnarray}%
As long as $k\leq N$ all the terms within the summations are zero, because
each of them contains a derivative of $G_{r}$ at most up to order $N$. The
first term generates a $G_{r}^{(k+1)}(\tau _{1})$. When $k+1\leq N$, this
term also vanishes. By assumption, the last term vanishes when $k+1\leq
N-(j-1)$. Since $j\geq 2$ this implies that all terms vanish when $k\leq N-j$%
. Overall, we have shown that
\begin{equation}
\partial _{\tau }^{k}|\psi _{j}^{\perp }(\tau _{1})\rangle =0,~\text{for}%
~1\leq j\leq N,~0\leq k\leq N-j.  \label{psi-perps}
\end{equation}%
Immediate corollaries of this result are as follows:
\begin{eqnarray}
|\psi _{j}^{\perp }(\tau _{1})\rangle &=&0,~1\leq j\leq N,
\label{proof-rem-2} \\
|\psi _{j}(\tau _{1})\rangle &\overset{(\ref{2.10})}{=}&f_{j}(\tau
_{1})|\Phi (\tau _{1})\rangle ,~1\leq j\leq N,  \label{cor-1} \\
\partial _{\tau }^{k}f_{j}(\tau _{1}) &\overset{(\ref{2.10}),(\ref{psi-perps}%
)}{=}&0,~1\leq j\leq N,~1\leq k\leq N-j.  \label{cor-2}
\end{eqnarray}%
This concludes the proof.
\end{proof}

%%%%%%%%%%%%%%%%%%%%%%%%%%%%%%%%%%%%%%%%%%%%%%%%%%%%%%%%%%%%%%%,
%\bibliographystyle{/home/lidar/revtex/prsty}
%\bibliography{/home/lidar/articles/bib}
%\bibliographystyle{/home/Daniel/revtex/prsty}
%\bibliography{/home/Daniel/articles/bib}

\end{document}